\documentclass[showpacs,twocolumn,pre]{revtex4}
\usepackage{psfrag,epsfig,amsfonts,amssymb,amsmath,graphicx,slashbox}
\usepackage{dcolumn}

\newcommand{\hr}{{\cal H}}
\newcommand{\ord}{{\cal O}}
\newcommand{\tr}{\mbox{Tr}}
\newcommand{\da}{\Delta_{\! A}}
\newcommand{\db}{\Delta_{\! B}}
\newcommand{\daf}{\Delta_{\! A-B({\bf b})}}
\newcommand{\dda}{\delta \! A}
\newcommand{\sa}{\sigma_{\! A}}
\newcommand{\mic}{\mathrm{mic}}

\newcommand{\RR}{{\mathbb R}}

\begin{document}

\title{Canonical Thermalization}

\author{Peter Reimann}
\affiliation{Universit\"at Bielefeld, Fakult\"at f\"ur Physik, 33615 Bielefeld, Germany}

\begin{abstract}
For quantum systems which are weakly coupled 
to a much ``bigger'' environment, thermalization 
of possibly far from equilibrium initial ensembles
is demonstrated: for sufficiently large times,
the ensemble is for all practical purposes
indistinguishable from a canonical density operator
under conditions which are satisfied under many, 
if not all, experimentally realistic conditions.
\end{abstract}


\maketitle

\section{Introduction}
\label{s1}
It is commonly accepted that Quantum Mechanics
describes the entire ``physical world''.
In particular, equilibrium Statistical Mechanics,
and possibly also its limitations,
should follow from Quantum Mechanics,
though a really satisfying ``derivation''
still does not seem available.
The main objective of the present paper
is such a derivation of the ``summit'' \cite{fey}
of equilibrium
Statistical Mechanics, namely the canonical
ensemble, from Quantum Mechanics 
in combination with certain, very weak 
assumptions regarding the preparation,
the observables, and the Hamiltonian
of the system.

A key issue in Statistical Physics 
is the lack of knowledge about many 
``details'' of any given ``real'' system.
In particular, the initial condition of the
typically $10^{23}$ particles is unknown apart 
from a few ``gross'' features, for instance 
the (approximate) total energy 
and a few additional macroscopic properties
in case of an non-equilibrium initial condition.
The standard way to deal with this incomplete 
knowledge is to consider a statistical ensemble 
(many repetitions of the ``same'' experiment,
formally described by a density operator)
instead of one particular ``realization'' 
of the experiment (formally described by either 
a pure or a mixed quantum mechanical state).
Much less appreciated is the fact that 
the concrete statistical ensemble for
any given ``real'' system is also 
largely unknown and would be at least as
hard to actually determine as the state
of any single realization.
It is therefore unavoidable to introduce 
certain postulates regarding this 
largely unknown statistical ensemble.
Their only justification is their 
plausibility, an admittedly 
subjective concept.
For instance, the present author
finds it {\em not} plausible to
assume for an isolated macroscopic system at 
equilibrium a microcanonical ensemble,
be it as a postulate {\em per se} or as 
a consequence of some other hypothesis
or principle (e.g. Jaynes principle):
Its key claim is that
if one would be able to actually determine the 
ensemble averaged occupation probabilities 
of all energy levels,
one would always find that all levels 
within some small energy-interval
are occupied with exactly equal probabilities 
and all other levels are not occupied 
at all.
Since level populations are constants of 
motion, the same properties must apply
to the possibly far from equilibrium 
initial ensemble.
This seems indeed very unlikely to be true
for any given ``real'' system, independently 
of any further ``details'' of the setup 
and the preparation.
For instructive numerical examples see e.g. 
Ref. \cite{rig08}.

Put differently, we find it plausible
that there exists a well-defined 
statistical ensemble describing the initial
state of any given ``real'' system.
But its details are unknown, probably 
very complicated, and already quite 
different even when the ``same'' experiment
is repeated in two different labs.
Hence the chance that the unknown ``true'' 
ensemble happens to agree with {\em any} 
particular ensemble we are postulating
is virtually zero.
Our present solution of this problem 
is not to assume
any specific form of the ``true'' initial
ensemble,
but only some very general ``gross'' features.
Namely, we will assume that the ensemble averaged
occupation probabilities of the energy levels 
can be locally (on the energy axis)
averaged in a well-defined manner.
We will provide good reasons to 
expect that real systems satisfy the 
assumption by closer inspection of
the preparation procedure at the
origin of the initial condition.

A second important deficit of knowledge
regards the appropriate Hamiltonian 
(and Hilbert space) of the system, 
generating the time evolution
of the initial ensemble. 
To begin with, we will take it for 
granted that the system 
can be treated as strictly closed
(isolated, autonomous), 
though in real systems small remnant interactions 
with the rest of the world are unavoidable.
The main reason is that standard Quantum 
Mechanics is only able to describe the 
evolution in time of closed systems.
Open systems, interacting and entangled 
with an environment, can be handled only
``indirectly'', by first including the entire
relevant environment of the open system 
into a closed supersystem,
then evolving the latter by standard 
Quantum Mechanics
and finally eliminating the environment 
again.
Likewise, we assume that after including 
all relevant perturbations 
``from outside'' into the considered 
system, it must be possible to theoretically
model it as strictly isolated from the 
rest of the world.
If this were not possible, the problem 
could only be treated by means of a 
generalization of Quantum Mechanics.
Many (often hidden) hypotheses of how
small external perturbations may modify
standard Quantum Mechanics have been
proposed.
We do not share the viewpoint 
that such extensions are a pivotal
point in the foundation of 
Statistical Mechanics.
Otherwise, an unavoidable consequence would be
that Statistical Mechanics does not ``work''
for strictly closed systems.
Also numerical evidence seems to
support our present viewpoint.

Taking for granted a closed system,
the ``right'' Hamiltonian (and Hilbert space)
is still far 
from obvious, and even the existence
of one particular ``true'' Hamiltonian for any 
given ``real'' system is questionable.
Our present way to deal with this problem is
to focus on ``generic'' Hamiltonians, while
leaving any further details unspecified.

The Quantum Mechanical time evolution 
generated by the Hamiltonian will not be
touched in any way, neither by heuristically
modify it to account for small remnant
external perturbations (see above), nor
by introducing any kind of approximation.
In other words, the well-known 
{\em time-inversion invariance of 
Quantum Mechanics is fully and rigorously 
maintained.}

Within the above general framework, the main subject 
of our present work is the long-time evolution 
of a largely arbitrary and possibly far from 
equilibrium initial ensemble.
Specifically, we address the two key claims
of Statistical Mechanics in this context:
(i) Equilibration: The ensemble
approaches a {\em stationary} long-time behavior.
(ii) Thermalization: Provided the ensemble 
exhibits a sharply peaked energy distribution, 
the long-time steady state is captured by 
(i.e. is experimentally indistinguishable from) 
the {\em microcanonical ensemble} corresponding to 
the given energy peak.
In particular, if the total closed system consist of
a subsystem of actual interest which is weakly 
coupled to a much ``bigger'' environment
(canonical setup), then
the steady state of the subsystem alone
(after eliminating/tracing out the environment)
is captured by the {\em canonical ensemble}.

A satisfactory derivation of these cornerstones
of Statistical Mechanics from Quantum Mechanics
is a long-standing and still unsolved fundamental problem
\cite{lud58,boc59,ergod,per84,jen85,deu91,sre96,tas98,das03,rei08,lin09},
and has recently attracted considerable renewed interest
in the context of (almost) integrable many-body quantum 
systems \cite{rig08,exp,rig07,cra08}.

In this work we address the questions
in how far and in which sense equilibration 
and thermalization can be derived
within our above specified general
 framework with 
particular emphasis on the 
canonical setup.
The main new results of our present paper
are established in Sects. \ref{s9} and \ref{s7},
demonstrating that in the long-time limit, 
the reduced state of the ``small'' system is 
for all practical purposes indistinguishable 
from a canonical ensemble.
More precisely, the true ensemble itself
remains time-dependent forever and thus 
quite different from the canonical
density operator, but the experimentally observable 
differences between the two ensembles
are unresolvably small for the overwhelming
majority of times.
The other sections of the present paper 
provide the pre-requisites needed in those
central Sects. \ref{s9} and \ref{s7}.
They mostly collect and unify,
but partially also extend
previously known material.
The final section \ref{s10}
contains the summary and outlook
with particular emphasis on closely related
recent works 
and on the subject of (almost) integrable 
many-body quantum systems, being at 
the focus of the present Special Issue.

\section{General Framework}
\label{s2}
According to Sect. 1, we consider an 
isolated system, incorporating all relevant 
parts of the environment (thermal baths, reservoirs etc.),
and modeled according to standard Quantum
Mechanics by an autonomous Hamiltonian $H$ 
on a (separable) Hilbert space $\hr$.
Specifically,
we focus on spatially finite (compact) systems
with a large (macroscopic) but finite particle number, 
corresponding of $f$ degrees of freedom with 
\begin{equation}
1\ll f <\infty \ .
\label{-2}
\end{equation}
As a consequence, all eigenvectors of $H$ represent
bound states and hence the spectrum of $H$ is discrete (quantized).
As usual, $|n\rangle$ ($n=0,\, 1,\, ...$) denote 
the (typically infinitely many) eigenvectors of $H$ 
and the corresponding
eigenvalues $E_n$ are assumed to be ordered,
\begin{equation}
E_0\leq E_1\leq E_2\leq ...
\label{-1}
\end{equation}
with a finite
ground state energy $E_0>-\infty$
\cite{rue69}.
In other words, the Hamiltonian can be written
as
\begin{equation}
H:=\sum_n E_n\, |n\rangle\langle n |
\label{0}
\end{equation}
where $\sum_n$ indicates a summation over 
all $n=0,\, 1,\, ...$.

For any function $g:\RR\to\RR$ we adopt the common definition
\begin{equation}
g(H):=\sum_n g(E_n)\, |n\rangle\langle n | \ .
\label{01}
\end{equation}
In the special case of a power series, $g(x)=\sum_k g_k x^k$, 
one readily sees that the (\ref{01}) reproduces
$\sum_k g_k H^k$, as it must be. 
But (\ref{01}) also covers more general functions 
$g(x)$.

{\em Degenerate energies} correspond to equality 
signs in (\ref{-1}).
Yet, sums like in (\ref{0}) and (\ref{01}) are 
meant to run over all $n$-values.

For any energy eigenvalue $E_n$, the projector 
onto the associated eigenspace of $H$ is 
given by
\begin{equation}
P_{E_n} := \sum_{E_m=E_n} |m\rangle\langle m|
\label{271} 
\end{equation}
where $\sum_{E_m=E_n}$ indicates a summation
over all $m$-values satisfying $E_m=E_n$.
It follows that $P_{E_m}=P_{E_n}$
for degenerate energies $E_m=E_n$.
Consequently, the identity operator 
\begin{equation}
1_{\hr}:=\sum_n |n\rangle\langle n|
\label{271a} 
\end{equation}
can be rewritten as
\begin{equation}
1_{\hr} = \sum_{E_n} P_{E_n}
\label{272} 
\end{equation}
where $\sum_{E_n}$ indicates a summation
over all {\em mutually different}
$E_n$-values, i.e. degenerate
energies only appear once in 
the sum.
Likewise, the Hamiltonian from (\ref{0})
can be rewritten as
\begin{equation}
H = \sum_{E_n} E_n\, P_{E_n} \ .
\label{273} 
\end{equation}

\subsection{Level Counting, Entropy, Temperature}
\label{s21}
In the following, we collect some ``well-know'' results 
of  ``elementary level counting''.
Though some details may be strictly speaking more 
subtle than we will say below, and possibly even 
not yet proven rigorously in sufficient generality, 
our present point of view is that these unsolved
issues of Statistical Physics are substantially 
less critical than those at the actual focus
of our present paper (equilibration and
thermalization, see Sect. \ref{s1}).
Differently speaking, we will henceforth
restrict ourselves to Hamiltonians $H$ which 
satisfy the properties given below and adopt 
the common opinion that they cover most cases
of practical relevance.

We emphasize that the entire present section
exclusively deals with properties of the 
energy eigenvalues of the Hamiltonian (\ref{0}).
In other words, we do not speak about system 
states at all, 
and hence the considerations of the
present section do not depend on
whether the system is in or out of equilibrium, 
since these are specifications of the system 
state, not of the system {\em per se}.

To begin with, the number of energy levels below
any given upper limit $E$ is defined as
\begin{equation}
\Omega(E):=\sum_n \Theta (E-E_n) \ ,
\label{1}
\end{equation}
where $\Theta(x):=\int_{-\infty}^x dy\, \delta(y)$ 
is the Heaviside step-function.
The entropy then follows as
\begin{equation}
S(E):=k_B\ln\Omega(E)
\label{2}
\end{equation}
where $k_B$ is Boltzmann's constant.
(The more common definition 
$S(E):=k_B[\ln\Omega(E)-\ln\Omega(E-\Delta E)]$
with a small but finite $\Delta E$ is well-known to 
be equivalent to (\ref{2}) but would be less convenient 
later on.)
Focusing on the most common case, 
the entropy is an extensive quantity.
For a system with $f$ degrees of freedom,
$S(E)/k_B$ is thus very roughly
speaking comparable in order of 
magnitude to $f$, 
\begin{equation}
S(E) / k_B = \ord(f) \ .
\label{3}
\end{equation}

It follows from (\ref{1}) and (\ref{3}) that
in macroscopic systems with $f=\ord(10^{23})$
degrees of freedom the energy levels are
unimaginably dense on any decent energy 
scale: for instance, within an energy interval 
of $10^{(-10^{20})}$ Joule there will still be
of the order of $10^{(10^{23})}$ levels.
On these exceedingly tiny scales, the step 
function $\Theta(x)$ appearing in (\ref{1}) is
considered to be actually ``washed out'',
so that $\Omega (E)$ becomes a reasonably
smooth function of $E$ with a well defined
derivative
\begin{equation}
\omega (E):=\Omega'(E) = \sum_n \delta (E-E_n) \ ,
\label{4}
\end{equation}
where also the delta-function $\delta(x)=\Theta'(x)$ is
considered as ``washed out'' over many
energy levels.
In other words, {\em $\omega (E)$ represents the
density of states}.
The correspondingly washed out entropy
(\ref{2}) gives rise to the usual 
definition of temperature,
\begin{equation}
T(E):=1/S'(E) \ .
\label{5}
\end{equation}
Combining (\ref{1}), (\ref{2}), and (\ref{4}), (\ref{5})
yields the useful relation
\begin{equation}
k_B T(E)=\Omega (E)/\omega(E) \ .
\label{6}
\end{equation}

We reiterate that in our present work, 
{\em entropy and temperature are by definition
given by (\ref{2}) and (\ref{5})}, 
and as such are for the
moment completely {\em independent of the
question of whether the considered 
system is at equilibrium or not}.
In particular, we did not establish any relation 
so far between the energy $E$ and the state of the 
system.

The following statements can be readily
verified for simple examples like the 
ideal gas.
More detailed estimates, which we omit here,
indicate that they in fact remain true
quite generally:
The actual dependence of the right hand side
of Eq. (\ref{3})
on $E$ is such  that the entropy approaches 
zero for $E\downarrow E_0$,
but for the rest (i.e. for macroscopic
values of $E-E_0$), the dependence on $E$ 
is comparably weak, essentially of 
logarithmic form.
Likewise, (\ref{5}) approaches zero for 
$E\downarrow E_0$.
Combining all these properties of $S(E)$ 
with (\ref{3}) we can conclude that
\begin{equation}
k_BT(E)=\ord\left(\frac{E-E_0}{f}\right)
\label{7}
\end{equation}
and hence
\begin{equation}
T(E+\Delta E)=T(E)\, \left[1+\ord\left(\frac{\Delta E}{E-E_0}\right)\right] \ .
\label{8}
\end{equation}
Note that if we did not avoid speaking about 
equilibrium states, then (\ref{7}) could also
be justified via the equipartition of energy
for an extensive systems with energy $E$
at equilibrium.

Finally, we can infer from (\ref{7}) and the derivative of this relation
in combination with (\ref{5}) that
\begin{equation}
-\frac{S''(E)}{k_B}=\frac{k_BT'(E)}{[k_BT(E)]^2} = \ord\left(\frac{f}{[E-E_0]^2}\right)
\ .
\label{8a}
\end{equation}

Similarly as in the first paragraph of this section,
we may take the alternative point of view that we
only consider model Hamiltonians $H$ which capture
the following common property of real systems 
reasonably well: 
If the macroscopic system energy $E$ is changed by 
an amount $dE$ then the microscopic kinetic energy 
per degree of freedom $k_BT(E)/2$ changes by an 
amount which is very roughly of the order of magnitude of 
$dE/f$, i.e.
\begin{equation}
k_BT'(E)=\ord (1/f) \ .
\label{8c}
\end{equation}
Since $T(E_0)=0$, integration of
(\ref{8c}) implies (\ref{7}) and (\ref{8a}).
Similarly, the third law $S(E_0)=0$ and (\ref{5})
yield upon another integration the relation 
(\ref{3}) and the logarithmic dependence of $S(E)$ 
upon $E-E_0$ mentioned above (\ref{7}).

It is well-known that all these relations may 
become problematic for extremely low temperatures.
Such cases are tacitly excluded from now on.

\subsection{System States and Dynamics}
\label{s23}
According to standard Quantum Mechanics,
the state of the system is at any
time instant $t$ given by a density
operator $\rho (t)$.
While we are mainly interested in
statistical ensembles (mixed states)
in this paper, it is nevertheless 
worth to point out that formally 
our considerations will also
{\em cover pure states as special case}.
The time evolution can be written as
\begin{equation}
\rho(t)=U_t\rho(0) U_t^\dagger
\label{25a} 
\end{equation}
with unitary propagator 
\begin{equation}
U_t:=\exp\{-iHt/\hbar\}=\sum_n \exp\{-iE_nt/\hbar\} |n\rangle\langle n |
\label{25b} 
\end{equation}
where we have exploited (\ref{01}) in the
last relation.
Denoting the matrix elements of $\rho(t)$ by 
\begin{equation}
\rho_{mn}(t) := \langle m|\rho(t)|n\rangle \ ,
\label{26} 
\end{equation}
Eqs. (\ref{25a}) and (\ref{25b}) yield
for an arbitrary initial condition $\rho(0)$ at 
time $t_0=0$ the result
\begin{equation}
\rho(t)=\sum_{mn} \rho_{mn}(0)\, e^{-i[E_m-E_n]t/\hbar}
\, |m\rangle\langle n| \ ,
\label{27}
\end{equation}
where $\sum_{mn}$ indicates a summation over all $m,n=0,1,2,...$.

\subsection{Level populations}
\label{s43}
Level populations are denoted by $p_n$ and refer to
the ensemble averaged occupation probabilities
of the energy eigenstates $|n\rangle$.
In other words, $p_n$ is the expectation value of the 
observable $|n\rangle\langle n|$. 
According to (\ref{26}) and (\ref{27}) it can be rewritten as
\begin{equation}
p_n := \tr\{ |n\rangle\langle n|\, \rho(t)  \}=\rho_{nn}(t)=\rho_{nn}(0)
\label{27x}
\end{equation}
independently of $t$.

Since every density operator $\rho (t)$ is, for any
value of $t$, non-negative and Hermitian, it
follows that $( \psi,\phi):=\langle\psi|\rho (t) +\epsilon|\phi\rangle$
satisfies the scalar product axioms for any $\epsilon>0$.
Hence, Cauchy-Schwarz's inequality applies,
i.e. $|(\psi,\phi)|^2\leq ( \phi,\phi)\, ( \psi,\psi)$.
In the limit $\epsilon\to 0$ we thus obtain
\begin{equation}
|\rho_{mn}(t)|^2\leq \rho_{mm}(t)\, \rho_{nn}(t) = p_n p_m \ .
\label{27a}
\end{equation}

Similarly as in (\ref{27x}), the ensemble averaged 
occupation probability $p_{E_n}$ of an energy
eigenvalue $E_n$ is given by the expectation value of 
the projector (\ref{271}) onto the corresponding 
eigenspace and can be rewritten as
\begin{equation}
p_{E_n} := \tr\{P_{E_n} \, \rho(t)\} = \sum_{E_m=E_n} \rho_{mm}(t)= \sum_{E_m=E_n} p_m \ ,
\label{278} 
\end{equation}
independently of $t$.
For convenience, also the $p_{E_n}$ will sometimes
be called level populations.

The obvious normalization conditions are
\begin{equation}
1 = \tr\rho(t)= \sum_{n} \rho_{nn} (t) =\sum_{n} p_{n} = \sum_{E_n} p_{E_n}\ .
\label{275} 
\end{equation}

Since $P_{E_n}\rho(0)P_{E_n}$ commutes with 
$H$ from (\ref{273}), we can without loss of 
generality, choose a basis in which both
operators are simultaneously diagonal.
In the following, {\em we always work with this
specific energy basis} due to its convenient
property that all the non-diagonal elements 
of $P_{E_n}\rho(0)P_{E_n}$ vanish,
\begin{equation}
\rho_{mn}(0) = 0 \ \mbox{if $m\not = n$ and $E_m=E_n$} \ .
\label{277} 
\end{equation}

Given any ensemble $\rho(t)$, 
let $\hr_+\subset\hr$ be 
the sub-Hilbert space spanned by those basis 
vectors $|n\rangle$ for which $p_{E_n} \not = 0$,
\begin{equation}
\hr_+ := \mbox{span}\{|n\rangle\, |\, p_{E_n}>0\} \ .
\label{27b}
\end{equation}
Exploiting (\ref{27a})
it follows that $\rho_{nm}(t)=0$ whenever
$\rho_{nn}(t)=0$ or $\rho_{mm}(t)=0$.
Denoting by $P_+$ the projector onto $\hr_+$
we thus can conclude that
\begin{equation}
\rho(t) := P_+\, \rho(t) \, P_+ \ .
\label{27c}
\end{equation}

\subsection{Observables}
\label{s24}
As usual, observables are represented by
Hermitian operators 
\begin{eqnarray}
A & = & \sum_{mn}A_{mn}\, |m\rangle\langle n|
\label{281x}
\\
A_{mn} &:= & \langle m|A| n\rangle
\label{282}
\end{eqnarray}
with expectation value
\begin{equation}
\langle A \rangle (t):= \tr\{\rho(t)A\} 
\label{28}
\end{equation}
and, without loss of generality, are 
assumed not to depend explicitly on time.

According to (\ref{27c}) it follows that
\begin{equation}
\langle A \rangle (t):= \tr\{\rho(t) A_+\} 
\label{28a}
\end{equation}
where $A_+$ is the projection/restriction of
$A$ onto $\hr_+$ from (\ref{27b}),
\begin{equation}
A_+ := P_+\, A \, P_+ \ .
\label{28b}
\end{equation}
In other words, only the sub-Hilbert space
$\hr_+$ and the projected/restricted 
observables (\ref{28b}) actually matter.
Hence we can from now on
{\em replace $\hr$ and $A$ by 
$\hr_+$ and $A_+$ whenever it will 
be convenient}.

\section{The problem of equilibration}
\label{s2a}
Generically, the statistical ensemble $\rho(t)$ 
is not stationary right from the 
beginning, in particular for an initial
condition $\rho(0)$ out of equilibrium.
But if the right hand side of (\ref{27})
depends on $t$ initially, it cannot 
approach for large $t$ any time-independent 
``equilibrium ensemble'' whatsoever.
In fact, any mixed state $\rho(t)$ returns 
arbitrarily
``near'' to its initial state $\rho(0)$ for certain, 
sufficiently large time-points $t$, 
and similarly for the expectation values 
(\ref{28}), as demonstrated for instance in
Appendix D of Ref. \cite{hob71}.
We emphasise, that these arbitrarily close
recurrences do not refer to pure states 
only (as in the classical Poincar\'e recurrences)
but rather to arbitrary statistical 
ensembles $\rho(t)$.

More specifically, consider any $\rho(t)$
which is not completely independent of $t$.
Then, according to (\ref{27}) there must 
exists at least one $\rho_{mn}(0)\not = 0$
with $\omega:=[E_n-E_m]/\hbar\not =0$.
In fact, one expects that one usually finds
pairs with $\omega$-values ranging from extremely
small to extremely large values on the scale of
1 Hz, thus including any 
experimentally ``reasonable'' frequency.
Focusing on the specific observable 
\begin{equation}
A=\hat A + \hat A^\dagger \ , \  \hat A:=|m\rangle\langle n| /\rho_{mn}(0)
\label{29}
\end{equation}
it readily follows from (\ref{27}) that 
\begin{equation}
\tr\{\rho(t)A\}=2\, \cos(\omega t) \ .
\label{30}
\end{equation}
In other words, the ensemble $\rho(t)$
exhibits permanent oscillations rather 
than equilibration, at least as far as 
the observable $A$ is concerned.

The main implication of the two previous
paragraphs is that equilibration,
as specified in Sect. \ref{s1}, cannot
be true and hence cannot be proven 
in full generality and rigor.
Put differently,
{\em Quantum Mechanics and equilibration
are strictly speaking incompatible}.
Equilibration can at most approximately
hold true for a restricted class of 
observables $A$ and initial conditions 
$\rho(0)$.
The main objective of our present work is to
show that, and in which sense this is indeed 
the case under rather weak restrictions
regarding observables and initial conditions.
Those are the subject of the next two Sections.

\section{Realistic Observables}
\label{s3}
The basic idea is that it is not necessary to
theoretically admit any arbitrary Hermitian 
operator $A$ as a possible observable 
\cite{realobs,lof,geo95,pop06}.
Rather it is sufficient to focus on
experimentally realistic observables in the 
following sense \cite{rei07}: 
Any observable $A$ must represent an experimental 
device with a {\em finite range} of possible 
outcomes of a measurement:
\begin{equation}
\da := 
\max_{\hr} \langle\psi|A|\psi\rangle
- \min_{\hr} \langle\psi|A|\psi\rangle 
= a_{max} - a_{min} \ ,
\label{31}
\end{equation}
where the maximization and minimization is over all 
normalized vectors $|\psi\rangle\in \hr$.
Accordingly,
$a_{max}$ and $a_{min}$ are the largest and 
smallest eigenvalues of $A$.

Moreover we require that this working range $\da$
of the device $A$ is limited to experimentally 
reasonable values compared to its 
{\em resolution limit $\dda$}.
All measurements known to the present author 
yield less than 20 relevant digits, i.e. 
$\da/\dda \leq 10^{20}$.
Maybe some day 100 or 1000 relevant digits will
become feasible, but it seems reasonable that a theory
which does not go very much beyond that will do.
We also remark that range and resolution are
specific to the given measurement device, but are
(practically) independent of the properties
(e.g. the size) of the observed system.

The above specified class of 
admissible observables clearly 
{\em includes any realistic 
measurement apparatus}.
Yet it turns out that
the class of observables which will be
admissible in our main results below
can still be substantially extended 
in two steps.

First, as said at the end of Sect.
\ref{s24}, we can replace the full Hilbert 
space $\hr$ by the sub-Hilbert space 
$\hr_+$ defined in (\ref{27b}). 
Accordingly, the full range 
$\da$ from (\ref{31}) can be replaced by
the reduced range
\begin{equation}
\da' := 
\max_{\hr_+} \langle\psi|A|\psi\rangle
- \min_{\hr_+} \langle\psi|A|\psi\rangle \ .
\label{31a}
\end{equation}
According to (\ref{27b}), $\hr_+$ is
at most as large as $\hr$.
However, in many cases the level populations (25)
may be safely negligible e.g. beyond some finite upper 
energy threshold, yielding a finite-dimensional
$\hr_+$, while $\hr$ is typically infinite 
dimensional.
Hence, the reduced range $\da'$ from (\ref{31a}) 
will be finite even for operators $A$ with an 
unbound spectrum on $\hr$, i.e. for
which the full range $\da$ from (\ref{31}) 
is infinite.

Second, we consider observables of the form
\begin{equation}
B({\bf b}):=\sum_n b_n\, |n\rangle\langle n|
\label{61b}
\end{equation}
with arbitrary real coefficients ${\bf b}:=(b_0,b_1,...)$.
In particular, we can conclude from (\ref{01}) that
arbitrary functions $g:\RR\to\RR$ of $H$ are of this form,
\begin{equation}
g(H)= B({\bf b}) \ \ \mbox{if}\ \ b_n:=g(E_n) \ .
\label{61x}
\end{equation}
As it will turn out, it is sufficient
to consider instead of $A$ in (\ref{31a})
any observable of the form $A-B({\bf b})$
with arbitrary coefficients ${\bf b}$.
As a consequence, $\da'$ from (\ref{31a}) 
can be replaced by
\begin{eqnarray}
\da'' & := & \min_{\bf b}\{
\max_{\hr_+} \langle\psi|A-B({\bf b})|\psi\rangle
\nonumber
\\
& & - \min_{\hr_+} \langle\psi|A-B({\bf b})|\psi\rangle\}
\label{31b}
\end{eqnarray}
where $\min_{\bf b}$ indicates a minimization 
over all real coefficients ${\bf b}:=(b_0,b_1,...)$.
In particular, it follows that
\begin{equation}
\da'' = 0 \ \mbox{if $A=B({\bf b})$ or $A=g(H)$}
\label{31c}
\end{equation}
for some set of coefficients ${\bf b}$ or
some function $g(x)$.

From (\ref{31}), (\ref{31a}), and (\ref{31b}) we 
see that
\begin{equation}
\da'' \leq \da' \leq  \da \ .
\label{31d}
\end{equation}
Rather than requiring that the full range-to-resolution
ratio does not exceed $10^{20}$, as discussed below 
(\ref{31}), it will be sufficient in our central result 
below to similarly limit the reduced ratio:
\begin{equation}
\da''/\dda \leq 10^{20} \ .
\label{32}
\end{equation}

\section{Realistic Initial Conditions}
\label{s4}
%
\subsection{Population density}
\label{s41}
Our key requirement with respect to the initial
condition $\rho(0)$ is that the concomitant
ensemble averaged level populations (\ref{27x}) 
can be written in the form
\begin{equation}
p_{n} = h (E_{n}) + \delta p_{n}
\label{706}
\end{equation}
with a smooth function $h(E)$ 
and ``unbiased fluctuations'' $\delta p_{n}$.
Physically, $h(E)$ thus represents a locally 
averaged level occupation probability, 
henceforth abbreviated as 
{\em population density}.

To be more precise with respect to (\ref{706}),
we recall that for a system with $f$ 
degrees of freedom, 
there are roughly $10^{\ord(f)}$ energy 
eigenvalues $E_n$ per
Joule, see Sect. \ref{s21}.
Assumption (\ref{706}) means that within 
any energy interval around some reference
energy $E>E_0$, which contains very many 
levels $E_{n}$, but which is still exceedingly 
small on any experimentally resolvable scale, 
the ensemble averaged level populations $p_n$
can be split into an approximately 
constant ``local'' average value $h(E)$ and 
``unbiased fluctuations'' $\delta p_{n}$,
i.e. the average over all
$\delta p_{n}$ belonging to this interval
around $E$ is negligibly small compared 
to $h(E)$ itself.
The key point is that $h(E)$ must be
independent of the exact choice of the
considered energy interval around $E$.

Comparable assumptions of well-defined 
``local averages'' are tacitly taken
for granted in many different 
physical contexts.
Likewise, we find it quite plausible
that the ensemble averaged level populations $p_n$,
though largely unknown (see Sect. \ref{s1}),
still satisfy our present assumption 
under experimentally realistic 
conditions.
Further arguments are provided in
Sect. \ref{s46a} below.

Finally, we emphasise once more that all these 
considerations concern {\em ensemble averaged} 
level populations, i.e. mean values over many
repetitions of an experiment, which the experimentalist
would denote as ``identical'' but which in fact
are very different on the microscopic level,
see Sect. \ref{s1}.

\subsection{System preparation}
\label{s46a}
The key idea is that the initial condition
is the result of a preparation process, 
during which the system was not yet isolated,
admitting conclusions about the initial 
condition itself.

The simplest case consists in a time-dependent
parametric change of the Hamiltonian during the
preparation phase.
More complex, but ultimately applying to every 
real experiment, is some type of contact with
the ``rest of the world'' prior to the actual
isolation of the system.

The first consequence is an entanglement with the
rest of the world during the preparation phase ($t<0$), 
implying that the reduced initial state (at $t=0$)
of the system 
(after tracing out the rest of the world) will be 
a mixed state even for a single realization of 
the experiment.
Already this reduction step brings along a certain
``randomization'' of the system level populations.
More importantly, there will unavoidably arise
some kind of time dependencies of the system
Hamiltonian during the preparatory period $t<0$,
the last of them being caused by
the actual shutting down of all 
connections with the rest of the world.
Such a time dependence of the system
Hamiltonian is known to generically
entail an approximately diffusive
``spreading'' of occupation probabilities
over neighboring energy levels
\cite{ediff}.
Since the levels are so exceedingly dense,
the diffusion will -- already during a 
very short time span and even for a 
very weak time-dependence of the 
Hamiltonian -- effectively lead to a
diffusive randomization of the
$p_n$'s in accordance with 
(\ref{706}).

In theoretical studies it
is quite common to generate the out of equilibrium
initial condition by means of a ``sudden''
(discontinuous) parametric change of the
Hamiltonian \cite{rig08,rig07,cra08},
called ``quantum quench''.
Such a procedure thus misses the
above mentioned diffusive ``spreading'' 
of the occupation probabilities.

\subsection{Energy density}
\label{s44}
The energy probability density, or
{\em energy density} for short, is defined as
\begin{eqnarray}
\rho(E):=\langle\delta(E-H)\rangle \ .
\label{710}
\end{eqnarray}
Accordingly, $\rho(E)\, dE$ quantifies the 
ensemble averaged probability to find a value between 
$E$ and $E+dE$ when measuring the energy of the system.

With (\ref{01}), (\ref{27}), (\ref{27x}), and (\ref{28}) it follows that
\begin{eqnarray}
\rho(E) = \sum_{n}p_{n}\ \delta(E-E_{n}) 
\label{712}
\end{eqnarray}
independently of $t$.
In the same spirit as around (\ref{4}) and in the
discussion of $h(E)$ below (\ref{706}), the delta-functions
in (\ref{710}) and (\ref{712}) are understood
to be ``washed out'' over many energy levels
in order to give rise to a well-defined, smooth
energy density. 
As a consequence one finds that \cite{rei07a}
\begin{eqnarray}
\rho(E) & = & h(E) \ \omega(E) \ .
\label{716}
\end{eqnarray}
While a detailed derivation of this relation is
provided in Appendix C , it is intuitively
quite obvious:
The probability $\rho(E)\, dE$ to encounter an energy 
between $E$ and  $E+dE$ is equal to the locally 
averaged population $h(E)$
of the energy levels multiplied by the local level 
density $\omega(E)$ times the interval length $dE$.

\subsection{Maximal level population}
\label{s45}
To get a feeling for the exotic orders
of magnitudes arising in the context of (\ref{706}),
let us assume that there are
exactly $10^{(10^{23})}$ equally spaced 
energy levels $E_n$ per Joule and that
our energy interval around 
$E$  has a length between 
$10^{-(10^{22})}$J and a few J.
Then, our interval contains 
at least roughly $10^{\ord(10^{23})}$ energy 
levels.
Assuming that $h(E)$ is approximately
constant within the interval, and zero 
outside, the normalization (\ref{275})
implies that 
$h(E)=10^{-\ord(10^{23})}$
within the interval.
Recalling that this is the local average
value of $p_n$, it seems quite reasonable
to assume that all the individual $p_n$
values do not exceed the range between
zero and $10^{(10^{22})}$
times the average value $h(E)$.
Otherwise, the average over the
$\delta p_n$'s would not be negligible 
compared to $h(E)=10^{-\ord(10^{23})}$ 
for every possible choice of the interval.

Returning to the general case, we can conclude 
that even if $h(E)$ varies very fast
on any experimentally realistic scales and
even if the energy levels are populated 
extremely unequally, we still expect that 
$\max_{n} p_n$ will be extremely small, 
typically
\begin{equation}
\max_{n} p_n=10^{-\ord(f)} \ .
\label{33}
\end{equation}
On this rather heuristic level,
(\ref{706}) thus implies (\ref{33}).
Intuitively it even seems plausible that 
the two conditions are more or less 
equivalent.

Next we remark that the mere existence
of the level density (\ref{4}) implicitly 
takes for granted that 
{\em the multiplicities of degenerate energies 
are not exceedingly large},
i.e. very much smaller than $10^{\ord(f)}$.
There can be little doubt that this assumption
will be fulfilled under all experimentally 
realistic conditions.
Under the very same assumption we can
infer from
(\ref{278}) and (\ref{33}) the very rough estimate
\begin{equation}
\max_{E_n} p_{E_n}=10^{-\ord(f)} \ .
\label{33a}
\end{equation}

Note that $p_{E_n}$ is the occupation probability
of $E_n$ from (\ref{278}), and as such does
not refer to any specific energy basis.
In contrast, (\ref{33}) is implicitly understood
with respect to the specific basis 
introduced above (\ref{277}).
This is the main advantage of (\ref{33a}) 
compared to (\ref{33}).

\subsection{Physical arguments}
\label{s46}
Besides those already discussed in Sect. \ref{s46a},
there are the following additional physical reasons 
to expect that
(\ref{33}) and (\ref{33a}) are fulfilled under
experimentally realistic circumstances.

First, the time-energy uncertainty relation
seems to prohibit for all practical purposes 
the determination of the system energy with a 
precision that would be necessary to populate
only a relatively small number of levels with 
appreciable probability so that
(\ref{33}) and (\ref{33a}) would be violated

Second, while an ideal energy measurement
would in principle allow us to prepare the system
at one specific energy eigenvalue, every real
(finite resolution)
measurement will result in appreciable
probabilities of very many levels.

\subsection{Example (\ref{29})}
\label{s47}
It is instructive to reconsider
our example from (\ref{29}), (\ref{30})
and see what happens to the
concomitant incompatibility with equilibration
in case we restrict ourselves 
to realistic observables and initial
conditions, satisfying
(\ref{32}) and (\ref{33}), respectively.
To begin with, one readily sees that 
the spectrum of $A$ from (\ref{29})
(within $\hr_+$)
consists of the two eigenvalues 
$a_{\pm}=\pm |\rho_{nm}(0)|^{-1}$,
and, in case dim$\hr_+>2$, of one further 
eigenvalue $a_0=0$.
With (\ref{27a}), (\ref{31}) and (\ref{31a})
we can conclude that 
\begin{equation}
\da=\da'=2|a_\pm|\geq 2/\max_{n} p_{n} \ .
\label{34}
\end{equation}
A somewhat more tedious calculation shows that
also $\da''$ from (\ref{31b}) coincides with $\da$
in our present example.
For experimentally realistic initial conditions
we can infer with (\ref{33})
that $\da\geq \ord ( 10^{f})$.
For macroscopic systems ($f\gg 1$)
it follows that the oscillations from (\ref{30})
are beyond any realistic experimental 
resolution limit $\dda$ 
according to (\ref{32}).
The same thing may alternatively also
be viewed as follows:
Any single (ideal) measurement process
always results
in one of the three outcomes
$a_+$, $a_-$, or $a_0$. 
Hence an infeasible number 
of repetitions is needed to resolve  
the order-one variations of the ensemble 
average (\ref{30}).

In short, while mathematically speaking
the observable (\ref{29}) indeed
leads to perpetual oscillations
(\ref{30}), such oscillations cannot
be resolved in practice for experimentally
realistic observables and 
initial conditions.

The above example also suggests that 
our assumptions of experimentally
realistic observables and initial conditions
(or some similar restrictions)
are almost unavoidable for taming the 
oscillations in (\ref{27})
and thus overcoming the concomitant 
incompatibility with the basic 
Statistical Mechanical
claim of equilibration,
see Sect. \ref{s2a}.

\section{Generic Hamiltonians}
\label{s5}
As detailed in Sect. \ref{s1}, the ``true''
Hamiltonian $H$ of a given system is usually
not known in detail.
Therefore, we assume that these
details are of ``generic'' character
in so far as the level counting properties 
from Sect. \ref{s21} are satisfied and 
energy differences $E_j-E_k$ and $E_n-E_m$ 
are never exactly equal apart from trivial 
cases.
More precisely, we require that

\begin{eqnarray}
& & \mbox{If $E_j\not = E_k$ and $E_m\not = E_n$}
\nonumber
\\
& & \mbox{then $E_j-E_k = E_n-E_m$} 
\nonumber
\\
& & \mbox{implies $E_j= E_n$ and $E_k = E_m$ .}
\label{52}
\end{eqnarray}

A condition similar to (\ref{52}) is 
well known under the names 
``non-resonance condition''
or ``non-degenerate energy gap condition'' 
and is considered to be satisfied by
generic Hamiltonians, see e.g. 
\cite{per84,sre96,tas98,lin09}, and, 
in particular, Sect. 3.2.1 in \cite{gol06} 
and references therein.
The essential intuitive argument is as follows:
Consider an arbitrary
``path'' $H(\lambda)$ in the ``space of all Hamiltonians'',
parameterized by $\lambda$.
In the absence of any special reasons like symmetries, 
it is quite plausible that every gap $E_n-E_m$ evolves as
a function of $\lambda$ somewhat differently than all 
the other gaps.
While we cannot exclude that two gaps may happen to
coincide for specific $\lambda$-values, these special
points are of measure zero.
In other words, Hamiltonians with degenerate energy gaps
are of measure zero compared to ``all'' Hamiltonians.

We remark that our present condition is weaker 
than the usual non-resonance condition 
\cite{per84,sre96,tas98,lin09,gol06} 
in so far as {\em (\ref{52}) still admits the
possibility of degenerate energy 
eigenvalues}.

\section{Equilibration for Isolated Systems}
\label{s6}
Being confident that the above discussed 
conditions
(\ref{32}), (\ref{33a}), and (\ref{52})
are fulfilled under many, if not all,
experimentally realistic conditions,
we henceforth take them for 
granted and turn to the question,
in how far they are sufficient to 
yield equilibration, i.e. a stationary 
long time behavior of the statistical 
ensemble (cf. Sect. \ref{s1}).

\subsection{Equilibrium ensemble}
\label{s61}
Given an arbitrary but fixed $\rho(0)$
evolving according to (\ref{27}),
we will see below
that the pertinent equilibrium ensemble 
is given by the density operator
(sometimes called 
the {\em generalized Gibbs ensemble})
\begin{equation}
\rho_{eq} := \overline{\rho(t)}
\label{64a}
\end{equation}
where the time average of an arbitrary function 
or operator $h(t)$ is defined as
\begin{equation}
\overline{h(t)}:=\lim_{T\to\infty}\frac{1}{T}\int_0^T dt\ h(t) \ .
\label{62}
\end{equation}
In other words, the equilibrium ensemble
$\rho_{eq}$ is the time averaged ``true'' 
ensemble $\rho(t)$.
As such, it is time-independent and
moreover inherits all the defining properties
of a genuine density operator from $\rho(t)$.
Namely, one readily sees that $\rho_{eq}$ 
is a non-negative, Hermitian operator of unit 
trace and satisfies the dynamics (\ref{27}).

Making use of the specific
basis introduced above (\ref{277}),
one can conclude -- as detailed in Appendix D  --
from (\ref{27x}) and (\ref{64a}) that
\begin{equation}
\rho_{eq}=\sum_n \rho_{nn}(0) \, |n\rangle\langle n|
= \sum_n p_n \, |n\rangle\langle n| \ .
\label{61}
\end{equation}
i.e., $\rho_{eq}$ amounts to the (time-independent) 
diagonal part of $\rho(t)$ from (\ref{27}).

Focusing on observables of the specific form (\ref{61b})
it follows with (\ref{27}) and (\ref{61}) that
\begin{equation}
\tr\{\rho(t)\, B({\bf b})\}=\tr\{\rho_{eq}\, B({\bf b})\} \ .
\label{61c}
\end{equation}
In particular, we can conclude with (\ref{61x}) that
\begin{equation}
\tr\{\rho(t)\, g(H)\}=\tr\{\rho_{eq}\, g(H)\} \ .
\label{61a}
\end{equation}
for arbitrary functions $g:\RR\to\RR$.

\subsection{Main result}
\label{s62}
It readily follows from (\ref{28}), (\ref{64a}), and (\ref{62}) that
\begin{equation}
\overline{\langle A\rangle (t)} = 
\tr\{ \rho_{eq}\, A\}
\label{64b}
\end{equation}
In other words, on the average over all times $t\geq 0$,
the ``true'' statistical ensemble $\rho(t)$ is 
indistinguishable from the equilibrium ensemble 
$\rho_{eq}$.

The natural next step is to consider the mean square deviation
\begin{equation}
\sa^2:=\overline{ [\langle A\rangle (t) - \overline{\langle A\rangle (t)}]^2} \ .
\label{65}
\end{equation}
The following relation is derived in Appendix D:
\begin{eqnarray}
\sa^2 \leq (\da'')^2\ \tr\{\rho_{eq}^2\} \ ,
\label{73}
\end{eqnarray}
where $\da''$ is defined in (\ref{31b}).
The last factor $\tr\{\rho_{eq}^2\}$ in (\ref{73})
is the so-called {\em purity of $\rho_{eq}$},
i.e. the purity of the time-independent part 
of $\rho(t)$,
but {\em not the purity of $\rho (t)$} itself.
In principle, $\rho (t)$ may even be a pure 
state (see above (\ref{25a})) with a 
purity of one, while the purity of the concomitant
$\rho_{eq}$ may still be as small as $10^{-\ord(f)}$
according to (\ref{33a}) and the relation
(\ref{281}) in Appendix B.

Observing that $\tr\{\rho_{eq}^2\}=\sum_n\rho_{nn}^2(0)$ 
according to (\ref{61}) and 
introducing relation
(\ref{281}) from Appendix B into (\ref{73}) we finally obtain 
\begin{eqnarray}
\sa^2 \leq (\da'')^2\ \max_{n} p_{E_n} \ ,
\label{74}
\end{eqnarray}
where $p_{E_n}$ is the occupation probability of $E_n$,
see (\ref{278}).

Considering $\tr\{\rho(t)A\}$ as a random variable,
generated by randomly sampling time points $t$
according to a uniform distribution on $[0,\infty)$,
the corresponding mean value and variance are
given by (\ref{64b}) and (\ref{65}).
The next step is to invoke Chebyshev's inequality
\cite{wik,hob71}, stating that for any random variable 
$x$ with average $\mu$ and variance $\sigma^2$
and any given $\kappa >0$,
the probability Prob$(|x-\mu|>\kappa)$
that $x$ deviates from $\mu$ by more than
$\kappa$ satisfies Prob$(|x-\mu|>\kappa) < (\sigma/\kappa)^{2}$.
In our present case we thus can conclude that 
\begin{eqnarray}
\mbox{Prob}\bigg(\big|\tr\{\rho (t)A \}-\tr\{\rho_{eq} A\}\big| \geq \dda  \bigg) 
\leq \left(\frac{\sa}{\dda}\right)^2 \ ,
\label{75}
\end{eqnarray}
where $\dda$ is the resolution limit of $A$,
see Sect. \ref{s3}.
With (\ref{74}) we arrive at the first main result of
our present paper:
\begin{eqnarray}
\mbox{Prob}\bigg(\!\big|\tr\{\rho (t)A \}-\tr\{\rho_{eq} A\}\big| \geq \dda\!  \bigg) 
\! \leq \! \left(\!\frac{\da''}{\dda}\!\right)^{\! \!2} \!\! \max_{n} p_{\! E_n}\, ,
\label{76}
\end{eqnarray}
where $p_{E_n}$ is the ensemble averaged
occupation probability of
the (possibly degenerate) energy eigenvalue $E_n$,
see (\ref{278}).
We recall that the only ingredients in deriving
this result were the (generalized) 
non-resonance condition (\ref{52}) 
and the assumption that the measurement range
$\da''$ from (\ref{31b}) is finite, cf. Sect \ref{s3}.
For the rest, (\ref{76}) is a completely general 
and rigorous relation, 
formally valid for any choice of $\dda>0$.
It generalizes the previously known result from 
\cite{rei08}, which did not admit degenerate
energy eigenvalues, nor the minimization
over arbitrary $B({\bf b})$ in (\ref{31b}).

\subsection{Discussion}
\label{s63}
For realistic initial conditions and generic
Hamiltonians we can take for granted the rough 
estimate (\ref{33a}), yielding with (\ref{76}) 
the result
\begin{eqnarray}
\mbox{Prob}\bigg(\!\big|\tr\{\rho (t)A \}-\tr\{\rho_{eq} A\}\big| \geq \dda\!  \bigg) 
\! \leq \! \left(\!\frac{\da''}{\dda}\!\right)^{\! \!2} 10^{-\ord(f)}\, .
\label{77}
\end{eqnarray}
Focusing on large systems
(\ref{-2}) we can conclude that for the overwhelming 
majority of times $t\geq 0$ the difference between 
$\tr\{\rho (t)A \}$ and $\tr\{\rho_{eq}A \}$ is way 
below the instrumental resolution limit $\dda$ 
for any experimentally realistic observable according
to (\ref{32}).
In other words, {\em the system looks exactly 
as if it were in the steady state ensemble $\rho_{eq}$ for 
the overwhelming majority of times $t\geq 0$},
though the ``true'' density operator 
$\rho(t)$ is actually quite
different, see Sect. \ref{s2a}.
This is the main result of our 
present paper regarding
the question of equilibration, 
see Sect. \ref{s1}.

Note that these conclusions do not
really require a macroscopic number
$f$ of degrees of freedom.
Put differently, our result also explains the
common numerical observation that
already quite small particle numbers 
often equilibrate and thermalize 
surprisingly well.

As promised in the introduction,
the derivation of (\ref{74})-(\ref{77}) is based on 
the exact Quantum Mechanical time evolution
(\ref{25a})-(\ref{27}) 
without any modification or approximation.
In other words, the full Quantum Mechanical
time-inversion invariance is still contained
in (\ref{77}).
In particular, (\ref{77}) is compatible with 
the recurrence property of $\tr\{\rho(t)A\}$
mentioned below (\ref{28b}), but implies 
that such excursions from  the
``apparent equilibrium state'' $\rho_{eq}$
must be exceedingly rare events.

Exactly the same ``apparent equilibration'' 
towards $\rho_{eq}$ emerges if one propagates 
$\rho(0)$ backward in time (keeping the 
system isolated also for $t<0$).
Along the entire real $t$-axis, 
an initial condition $\rho(0)$ far from 
equilibrium thus closely resembles one of the
above mentioned rare excursions, except that
the location of this excursion is on purpose
chosen as the time-origin.

In other words, {\em Quantum Mechanical 
time inversion invariance is maintained,
but when starting out of equilibrium, 
an ``apparent time arrow'' emerges with
extremely high fidelity}.

Note that any single excursion of $\tr\{\rho(t)A\}$
from the ``apparent equilibrium value'' $\tr\{\rho_{eq}A \}$
is a priori not expected to exhibit any special
symmetry with respect to time inversion.
Only the probabilistic properties of an ensemble
of such excursions are, in the absence of
magnetic fields, expected to satisfy a microreversibility
or detailed balance type of symmetry with respect
to time inversion.
Note that these considerations apply both to
``small'' and ``large'' excursions.

While (\ref{77}) provides a bound for 
the {\em relative} amount
of time the system exhibits notable
deviations from equilibrium, 
the typical duration of one given 
excursion, or equivalently, the characteristic
relaxation time of an out of equilibrium initial
condition $\rho(0)$ remains unspecified.
Note that Statistical Mechanics itself also
makes not statements in this respect.
Hence, it is justified to omit
them within a foundation of Statistical 
Mechanics.
However, we remark that since our assumptions
on initial condition and Hamiltonian
were very weak and we kept the
exact Quantum Mechanical time evolution
(\ref{25a})-(\ref{27}), we expect that
the actual relaxation time will be close to
that of the real system we are modeling,
provided this modeling is not too bad.
Since one can easily imagine real systems
with arbitrarily large or small 
relaxation times,
any further quantification of the
relaxation process inevitably would require 
a considerably more detailed specification 
of the Hamiltonian $H$,
the initial state $\rho(0)$, and the
observable $A$.
Thus, our main result (\ref{77}) 
may well be already quite close to 
``the maximum one can say in full generality''.

As mentioned above (\ref{25a}) and
below (\ref{73}), in principle 
{\em $\rho (t)$ may even be a pure state} 
of the form $|\psi(t)\rangle\langle\psi(t)|$.
In this case, the occupation probabilities
$p_{E_n}$ of the energy eigenvalues $E_n$
appearing in (\ref{76}) can be rewritten
according to (\ref{278}) as
\begin{equation}
p_{E_n}= \sum_{E_m=E_n} |\langle\psi(t)|m\rangle|^2 \ .
\label{77a}
\end{equation}
As long as all these occupation probabilities
are small, e.g. satisfying the rough estimate
(\ref{33a}), the above relation (\ref{77}) and
the subsequent discussion still remain valid 
for pure states 
$\rho (t) =|\psi(t)\rangle\langle\psi(t)|$.

If the system is prepared in a
pure energy eigenstate, i.e. 
$\rho(t)=\rho(0)=|n\rangle\langle n|$ then 
$\da'' =0$ for arbitrary $A$ according to (\ref{27b}) 
and (\ref{31b}).
In other words, (\ref{77}) still
represents a tight upper estimate
in this case, which in fact represents 
``the opposite extreme'' compared to the 
property (\ref{33}) or (\ref{33a})
of experimentally realistic initial 
conditions.

Likewise, for observables of the form
(39) or of the form $g(H)$
with arbitrary $g$ we have $\da''=0$
in (\ref{77}) according to (\ref{31c}).

We close with the following conceptual
remarks regarding the notion of 
``experimentally realistic observables 
and initial conditions''.
In Sects. \ref{s3} and \ref{s4} we have
specified certain properties which we are
proposing to be {\em necessary} for
observables or initial conditions to be
considered as ``experimentally realistic''.
But these properties are not meant to 
be {\em sufficient}.
While establishing such exact (necessary and sufficient)
conditions is not the subject of our present
work, we provide some simple example
to illustrate our point:
For any $A$ and any fixed $\tau$, 
the observable $B:=U_\tau A U_\tau^\dagger$
satisfies $\tr\{\rho(t)B\}=\tr\{\rho(t-\tau) A\}$
according to (\ref{25a}).
Whenever $A$ was realistic according 
to the criterion from Sect. \ref{s3}, the
same applies to $B$ ($\db=\da$).
According to (\ref{28}), $B$ imitates (for {\em any} $\rho(t)$) 
the behavior of $A$ with a time delay of $\tau$.
If $\rho(0)$ is a far from equilibrium initial condition and
$\tau$ exceeds the relaxation time for the time inverted 
dynamics, the observable
$B$ thus initially behaves as if the system were
already equilibrated but then all of a sudden
undergoes an excursion as if the system would
transiently move very far from equilibrium.
Turning to negative $\tau$ values, $B$ would
represent a device which can ``look''
-- in principle arbitrarily far -- 
into the future.
There can be little doubt that such 
observables are not ``realistic''.
Likewise, for any given
$\rho(t)$ satisfying the 
Quantum Mechanical time evolution (\ref{25a}),
a hypothetical initial condition of the form
$\tilde\rho_\tau(0):=\rho(-\tau)$ produces
analogous ``unrealistic'' phenomena
while being ``experimentally realistic''
according to Sect. \ref{s4} whenever
$\rho(0)$ was so
(the level populations of $\rho(0)$ and of
$\tilde\rho_\tau(0)$ are equal).
To identify suitable criteria for sorting out
such pathologies is a very subtle task,
especially in view of the fact that
so-called spin-echo experiments seem
indeed to be able to realize such
initial conditions $\tilde\rho_\tau(0)$
to some extent \cite{skl93}.

\section{The Problem of Thermalization}
\label{s8}
According to (\ref{61}) and the discussion
below (\ref{77}), expectation values (\ref{28})
become practically indistinguishable from
\begin{equation}
\tr\{\rho_{eq} A\} =\sum p_n A_{nn} 
\label{101}
\end{equation}
after initial transients have died out.
In this respect the problem of equilibration
raised in Sect. \ref{s1} can be considered 
as settled and we henceforth can focus on 
(\ref{101}).

Turning to the issue of thermalization,
the key question is thus,
in how far the equilibrium expectation value
of $A$ from (\ref{101}) 
is in agreement with that predicted by the
microcanonical ensemble, namely
\begin{equation}
\tr\{\rho^{\mic} A\} =\sum p^{\mic}_{n} A_{nn} \ ,
\label{102}
\end{equation}
where the level populations $p^{\mic}_{n}$
are equal to a normalization constant if
$E_n$ is contained within a small 
energy interval 
\begin{equation}
I:=[E-\Delta E, E] 
\label{103}
\end{equation}
and zero otherwise \cite{lldiu}.

In case (\ref{101}) and 
(\ref{102}) yield measurable 
differences for experimentally 
realistic $\rho(0)$ and $A$, the 
``purely Quantum Mechanical'' prediction (\ref{101})
is commonly considered as ``more fundamental''
\cite{rig08,rig07,cra08}.
From this point of view, {\em our derivation of 
equilibrium Statistical Mechanics is 
complete, provided the latter is valid itself}.

What are these validity conditions, beyond
which the microcanonical formalism of
equilibrium Statistical Mechanics may 
break down?

A first well known validity condition for 
the microcanonical formalism
is, as said below (\ref{102}), that only
$E_n$ within a small energy interval 
(\ref{103}) have a non-vanishing occupation 
probability.
More generally, as already mentioned in 
Sect. \ref{s1}, in equilibrium Statistical
Mechanics it is taken for granted that the
system energy is fixed up to unavoidable
experimental uncertainties.
On the other hand,
realistic initial conditions according to 
Sect. \ref{s4} in particular require 
that this energy uncertainty is much larger than
$10^{-\ord(f)}$ Joule, which is obviously
always fulfilled in practice, but we never
introduced or exploited any type of 
{\em upper limit} for this uncertainty so far, i.e.
{\em the energy uncertainty may still be 
arbitrarily large in (\ref{77}) and (\ref{101})}.
In other words, for large energy uncertainties,
our key relation (\ref{101}) remains valid, 
while equilibrium Statistical Mechanics is 
likely to become invalid.
This is clearly a not at all surprising case
of disagreement between (\ref{101}) 
and (\ref{102}) .

To avoid such ``almost trivial'' cases, 
we henceforth take for granted that
the system energy is known up to an uncertainty 
$\Delta E$ which is as small as possible, 
but still experimentally realistic, cf. 
Sect.\ref{s4}.

A second (often tacit) validity condition 
of the microcanonical formalism is that
the expectation values (\ref{102})
are required/assumed to be (practically) 
independent of the exact choice of of the 
interval $I$ in (\ref{103}), 
i.e. of its upper limit $E$ and its
width $\Delta E$.
But essentially this means nothing else 
than:
\begin{equation}
\mbox{In (\ref{101}) the details of $p_{n}$
are largely irrelevant.}
\label{p}
\end{equation}

The same conclusion (\ref{p}) follows from the 
equivalence of the microcanonical and canonical 
ensembles (for all energies $E$), considered 
as a self-consistency condition for 
equilibrium Statistical Mechanics \cite{lof,geo95}.

Clearly, given property (\ref{p}) holds, the expectation
values (\ref{101}) and (\ref{102}) are
indeed practically indistinguishable.

Our first remark regarding property (\ref{p}) 
itself, is that no experimentalist can control 
the populations $p_n$ of the unimaginably dense 
energy levels $E_n$, apart from the very gross
fact that they are ``mainly concentrated within the
interval $I$ from (\ref{103})''.
If the details would matter, not only 
equilibrium Statistical Mechanics would 
break down, but also reproducing measurements,
in particular in different labs, would 
be largely impossible, see also the discussion
in Sects. \ref{s1} and \ref{s4}.
Second, one can readily construct 
observables and initial conditions, being
experimentally realistic according to our
definitions in Sects. \ref{s3} and \ref{s4}
but still violating (\ref{p}).
The fact that equilibrium Statistical Mechanics
is known to have an extremely wide experimental 
applicability implies that our so far notion of
``experimentally realistic'' is still
too general (see also at the end of Sect. \ref{s63}).

The simplest way to guarantee property (\ref{p}) 
seems to require/assume 
that the expectation values $A_{nn}=\langle n|A |n\rangle$
hardly vary within any small energy interval 
of the form (\ref{103}).
This is similar in spirit to classical 
coarse graining, and, in fact, 
is part of a common 
conjecture about the semiclassical 
behavior of fully chaotic classical 
systems \cite{arg}.
In particular, negligible variations of 
$A_{nn}$ for close by $n$-values
imply Serdicki's ``eigenstate
thermalization hypothesis'' \cite{sre94} 
(anticipated in \cite{jen85} and
revisited in Ref. \cite{rig08}), implying 
that each individual energy eigenstate 
$|n\rangle$ behaves like the equilibrium 
ensemble.

An alternative way to guarantee property (\ref{p})
follows from the argument by Peres \cite{per84}
that even if the $A_{nn}$ may notably vary with $n$,
the immense number of relevant summands in 
(\ref{101}) may
-- for ``typical'' $A$ and $\rho(0)$ -- 
lead to a kind of statistical 
averaging effect and thus a largely 
$\rho(0)$-independent overall value 
of the sum.

Numerically, the validity and possible failure
of such conjectures and of property
(\ref{p}) itself have been exemplified e.g. in 
\cite{rig08,rig07,fei84,jen85}.
While the details -- in particular the 
role of  ``more basic'' system properties 
like ``ergodicity'' and ``(non-)integrability'' --
are still not very well understood 
\cite{rig08,ergod,per84,arg,wei92,rou09},
``equilibration'' in agreement with 
(\ref{101}) was seen numerically 
in all known cases.

\section{Canonical Setup}
\label{s9}
The objective in the remainder of the paper 
is to establish thermalization without making 
use of the unproven property (\ref{p}).
While the general case seems extremely difficult
to tackle, as discussed above, we focus on the
most important special case, namely
the so-called canonical setup: an isolated
compound system, consisting of a (sub-)system 
which is weakly coupled to a ``heat bath''.

In analogy to Sect. \ref{s2},
the starting point is a system (subsystem, central system, system of actual interest, index ``S'')
with $f_S$ degrees of freedom, Hamiltonian $H_S$, and Hilbert space $\hr_S$
together with an environment (e.g. a heat bath, index ``B'')
with $f_B$ degrees of freedom, Hamiltonian $H_B$, and Hilbert space $\hr_B$.
As usual, the environment is assumed to be macroscopic and much ``bigger''
than the system, i.e.
\begin{equation}
f_B\gg f_S \ .
\label{9a}
\end{equation}
The {\em system $S$ may or may not be macroscopic}, i.e. $f_S$ may but needs
not be a large number (in the range of $10^{23}$).
On the other hand $f_B$ is of the order of $10^{23}$
or even larger (e.g. if $f_S$ is already of this order).
The system-plus-environment compound (total system, supersystem, no index)
thus has
\begin{equation}
f = f_S+f_B
\label{9}
\end{equation}
degrees of freedom and ``lives'' in the product space 
\begin{equation}
\hr:=\hr_S\otimes\hr_B \ .
\label{10}
\end{equation}

The contact (coupling) between system and environment is described
by an interaction Hamiltonian $H_{int}\, :\, \hr\to\hr$ and
a ``coupling strength'' $\lambda$, resulting in a total
Hamiltonian of the form
\begin{equation}
H(\lambda)=H_S\otimes 1_{\hr_B} + 1_{\hr_S}\otimes H_B +\lambda\, H_{int}\ ,
\label{11}
\end{equation}
where $1_{\hr_S}$ indicates the identity on $\hr_S$, 
and similarly for $1_{\hr_B}$.

We will mainly be interested in observables which 
only concern system properties, i.e. which are of the form
\begin{equation}
A=A^S\otimes 1_{\hr_B} \ .
\label{n10}
\end{equation}

Within this general framework, we take for granted 
that all conditions for equilibration of the isolated 
system-plus-bath compound
in the sense of Sect. \ref{s6} are fulfilled, 
i.e. the observables (\ref{n10}) satisfy (\ref{32}), 
the ensemble averaged energy level populations satisfy (\ref{33a}),
and the Hamiltonian (\ref{11}) satisfies the generalized
non-resonance condition (\ref{52}).
As expected and demonstrated in detail later, 
these requirements in particular rule out $\lambda=0$.
A further requirement is the subject of the next subsection.

\subsection{Weak coupling condition}
\label{s92}
Some kind of weak coupling assumption is an 
indispensable (though often tacit) prerequisite 
of the canonical formalism.
The simplest possibility would be to require 
that $\lambda$ in (\ref{11}) is so small that
the eigenvectors $|n\rangle$ and eigenvalues $E_n$ 
of $H(\lambda)$ deviate only very little
from those of $H(0)$.
However, according to ordinary perturbation theory,
these deviations will be governed by terms of the form
$\lambda\langle m|H_{int}|n\rangle/[E_m-E_n]$.
Since some $E_m-E_n$ are of the order of $10^{-\ord(f)}$ Joule
(see below (\ref{3})), the admissible $\lambda$-values 
would be so small that no realistic model 
would satisfy the condition.

For this reason, we henceforth focus on 
system-plus-bath compounds (\ref{11}) 
which satisfy the following ``operational''
weak coupling condition:
{\em After equilibration of the system-plus-bath 
compound, a reversible (adiabatically slow) 
decoupling of the system from the bath
does not lead to any experimentally 
resolvable changes.}

As far as expectation values of system 
observables $A$ are concerned, 
we recall that the ``true'', time-dependent
expectation values $\langle A\rangle(t)$ 
may, even after equilibration, 
still exhibit quite notable ``excursions'' 
at exceedingly rare time points $t$ 
(cf. Sect. \ref{s6}).
In this case, the above weak coupling condition 
tacitly refers to time-averaged 
expectation values.

The quantities, for which the weak coupling
condition will actually be taken for granted 
later on, are observables of the form (\ref{n10})
and  the energy density (\ref{710})
of the total system-plus-bath compound.

Physically, there can be little doubt that 
most real systems in contact with a heat bath
satisfy the above weak coupling condition.
Hence, the same is expected for 
``realistic models'' of such systems.
Yet, in view of the perturbation theoretical 
considerations above, a mathematical proof 
for any given model seems extremely 
difficult.
In fact, the condition concerns not only 
the Hamiltonian (\ref{11}) 
but simultaneously the observables (\ref{n10}) and 
initial conditions $\rho (0)$ (cf. Sect. \ref{s4})
of the system-plus-bath compound.
The nature and difficulty of the problem may become more
evident by considering a particularly ``simple''
special case, namely observables of the form 
(\ref{n10}) and canonical density operators.
In this special case, we may consider the relation
\begin{equation}
\tr\left\{\frac{e^{-H(\lambda)/k_BT}}{Z(\lambda)}\, A\right\}
=
\tr_S \left\{\frac{e^{-H_S/k_BT}}{Z_S}\, A^S\right\}  \ ,
\label{n15}
\end{equation}
with $A$ from (\ref{n10}), $H(\lambda)$ from (\ref{11}),
$\tr$ and $\tr_S$ indicating the traces over the
Hilbert spaces $\hr$ and $\hr_S$ in (\ref{10}), 
respectively, and $Z(\lambda)$ and $Z_S$ representing standart
partition sums, normalizing the respective density 
operators.
This relation (\ref{n15}) is an elementary identity for 
$\lambda=0$, is usually considered as ``obvious'' 
for ``weak coupling'', but to the best of the present
authors knowledge 
is unproven (and probably wrong and thus unprovable without 
strong extra assumptions on $A^S$) for small but 
still experimentally realistic coupling strengths 
$\lambda$.

Since the solution of these long standing and very subtle
problems is not the actual main theme of our present work,
we adopt the standpoint that the above weak coupling
condition is an implicit additional requirement 
regarding experimentally realistic 
Hamiltonians, observables, 
and initial conditions
on top of the previous requirements
(\ref{32}), (\ref{33a}), (\ref{52}).
In all what follows, we focus on systems
which satisfy all those requirements, being 
confident that they include the majority of
experimentally realistic models.

\subsection{Main implications of weak coupling}
\label{s93}
The ``true'' time-averaged expectation values
are, according to (\ref{64b}), given by
\begin{equation}
\overline{\langle A \rangle(t)} =
\tr\{\rho_{eq}(\lambda) A\} \ ,
\label{n20}
\end{equation}
where the argument $\lambda$ of $\rho_{eq}$
has been added to remind us of the fact
that we are dealing with the given, 
``true'' Hamiltonian (73) with a ``small'' 
but non-vanishing coupling strength $\lambda$.
Likewise, the ``true'' density operator from
(\ref{27}) is now rewritten in the form
\begin{equation}
\rho(t)=\sum_{mn} \rho_{mn}(0)\, e^{-i\omega_{mn}(\lambda) t}
\, 
D_{mn}(\lambda)
\label{n30}
\end{equation}
where $\omega_{mn}(\lambda):=[E_m(\lambda)-E_n(\lambda)]/\hbar$,
$D_{mn}(\lambda):=|m(\lambda)\rangle\langle n(\lambda)|$, 
and where $|n(\lambda)\rangle$ and $E_n(\lambda)$ refer to
the eigenvectors and eigenvalues of the 
Hamiltonian (73) for an arbitrary but 
fixed $\lambda$-value.

As soon as $\lambda$ in (\ref{11})
starts to change in the course 
of time, the density operator is no longer given
by (\ref{27}) but rather follows from the 
Liouville-von Neumann equation
\begin{equation}
i\hbar \dot \rho(t)=[H(\lambda(t)),\rho(t)] \ .
\label{n40}
\end{equation}
While an explicit solution for general protocols $\lambda(t)$
is hopeless, in the special case of adiabatically slow
(quasi-static) parameter changes, the Adiabatic Theorem
can be invoked to yield 
\begin{eqnarray}
\rho(t)=\sum_{mn} \rho_{mn}(0)\, e^{-i\int_0^t \omega_{mn}(\lambda(s))\, ds}
\, 
D_{mn}(\lambda(t)) \ ,
\label{n50}
\end{eqnarray}
where we have tacitly restricted ourselves to the
simplest and most relevant case of non-degenerate 
energy levels, see also
Appendix E.

The adiabatically slow decoupling process appearing in the
weak coupling condition (Sect. \ref{s92}) means that $\lambda(t)$
in (\ref{n50})
is given during a large but finite initial time span by
the ``true'', finite coupling strength, then adiabatically slowly
changes to the value zero, and afterwards remains zero for all
later times $t$ until infinity.
Consequently, the time average of the density operator in (\ref{n50})
is governed by the infinitely long time period with $\lambda(t)=0$,
i.e.
\begin{equation}
\overline{\rho(t)}=\sum_n\rho_{nn}(0) \, 
|n(0)\rangle\langle n(0)|
\label{n60}
\end{equation}
Invoking the weak coupling condition from Sect. \ref{s92} it follows that
the ``true'' time-averaged expectation values (\ref{n20}) are practically 
indistinguishable from those obtained with the help of (\ref{n60}), i.e.
\begin{equation}
\overline{\langle A \rangle (t)} = \sum_n p_n\,  \langle n(0)|A|  n(0)\rangle
\label{n70}
\end{equation}
where $p_n:=\rho_{nn}(0)=
\langle n(\lambda(0))|\rho(0)|n(\lambda(0))\rangle$ 
are the level populations of the 
``true'' system at time $t=0$, cf. (\ref{27x}).
In other words,
{\em though the true system-bath coupling strength
$\lambda$ is finite, we can formally work in the zero 
coupling limit as far as time-averaged expectation 
values are concerned}.

A well-known further consequence of the Adiabatic Theorem 
(\ref{n50}) is the time-independence of the level populations 
$\langle n(\lambda(t))|\rho(t)|n(\lambda(t))\rangle$ which
can thus be identified with $p_n=\rho_{nn}(0)$ for all times $t$.
Considering $\lambda$ rather than $t$ as independent variable,
we may equivalently say that the $p_n$ are $\lambda$-independent.
It follows that in the relation 
$p_n=h(E_n(\lambda))+\delta p_n$ from (\ref{706}), 
the right hand side must be $\lambda$-independent
in the same sense.
By locally averaging over many $n$-values
(see below (\ref{706}) and \cite{f1})
we can conclude that also 
$h(E_n(\lambda))$ must be $\lambda$-independent.
In view of the $\lambda$-dependence of $E_n(\lambda)$ it
follows that also the function $h(E)$ generally must acquire a 
dependence on $\lambda$.
Indicating this fact by adding an index $\lambda$ 
to $h(E)$ we can conclude that
\begin{equation}
h_\lambda(E_n(\lambda))=h_0(E_n(0))
\label{n80}
\end{equation}
for all $n$, and that the relation (\ref{706}) now takes the form
\begin{equation}
p_n=h_\lambda (E_n(\lambda))+\delta p_n 
\label{n85}
\end{equation}
with $\lambda$-independent $p_n$ and $\delta p_n$.
Likewise, the density of states (\ref{4}) 
and the energy density (\ref{712}) acquire a 
$\lambda$-dependence and thus are 
now denoted by $\omega_\lambda(E)$ and 
$\rho_\lambda(E)$, respectively.
Similarly as in (\ref{716}) one can conclude 
\cite{f1} that
\begin{equation}
\rho_0(E)=h_0(E)\, \omega_0(E) \ .
\label{n90}
\end{equation}

Finally, the weak coupling condition as discussed in
Sect. \ref{s92} includes the practical indistinguishability 
of the ``true'' energy density from the energy density after
decoupling the system from the bath, i.e.
\begin{equation}
\rho_\lambda(E)=\rho_0(E)
\label{n100}
\end{equation}

\subsection{Zero coupling limit}
\label{s91}
We consider the Hamiltonian (\ref{11}) in the limit
of vanishing coupling strength $\lambda$.
Denoting by $|n\rangle_S$ and $E_n^S$ the eigenvectors 
and eigenvalues of $H_S$ and
by $|m\rangle_B$ and $E_m^B$ those of $H_B$,
those of (\ref{11}) with $\lambda=0$ follow as
\begin{eqnarray}
|nm\rangle & := & |n\rangle_S|m\rangle_B \ ,
\label{12a}
\\
E_{nm} & := & E_n^S + E_m^B \ .
\label{12b}
\end{eqnarray}

\subsubsection{Violation of the non-resonance condition}
\label{s912}
The fact that systems consisting of non-interacting 
sub-systems require special attention with respect 
to the generalized non-resonance condition
(\ref{52}) has first been noticed in
Ref. \cite{lin09}.

Taking into account that the original indices 
in (\ref{52}) now become double indices 
according to (\ref{12b}), condition (\ref{52})
takes the modified form
\begin{eqnarray}
& & \mbox{If $E_{j_1j_2}\not = E_{k_1k_2}$ and $E_{m_1m_2}\not = E_{n_1n_2}$}
\nonumber
\\
& & \mbox{then $E_{j_1j_2}-E_{k_1k_2}=E_{n_1n_2}-E_{m_1m_2}$}
\\
\nonumber
& & \mbox{implies $E_{j_1j_2}= E_{n_1n_2}$ and $E_{k_1k_2} = E_{m_1m_2}$ .}
\label{513}
\end{eqnarray}
One readily sees that this condition is violated
by considering the following specific choice
\cite{lin09}:
$j_1=j_2=k_1=n_2=:n$ and $k_2=m_1=m_2=n_1=:m$.
In the generic case, it will be possible to
find indices $n$ and $m$ so that all four 
energies $E_{nn}$, $E_{nm}$, $E_{mn}$, $E_{mm}$ 
appearing in (88) in are different.
With (\ref{12b}) it follows that
$E_{nn}-E_{nm}=E_{mn}-E_{mm}$.
In other words, condition (88)
is violated.

\subsubsection{Exploiting the product basis}
\label{s913}
The purpose of this subsection is to rewrite the pertinent relations
from Sect. \ref{s93} in terms of the product energy basis (\ref{12a})
and the corresponding energy eigenvalues (\ref{12b}) in the zero 
coupling limit. 

Essentially, we are just relabeling all 
eigenvectors and eigenvalues in a way which will turn out
particularly convenient later on.
Namely, all the single labels $n(\lambda)$ are now replaced by
double lables $nm(\lambda)$ with the additional convention 
that the argument $\lambda$ will be omitted in the case $\lambda=0$,
in agreement with the notation in Sect. \ref{s912}.
Note that only in the zero coupling limit does the
first of the two indices in $nm$ refer to
the system and the second to the bath, but not any more 
for $\lambda\not = 0$.

Specifically, $|n(0)\rangle$ and $E_n(0)$ from Sect. \ref{s93} are 
now denoted as $|nm\rangle$ and $E_{nm}$ and satisfy 
(\ref{12a}) and (\ref{12b}).
Likewise, $|n(\lambda)\rangle$ and $E_n(\lambda)$ 
are now denoted as $|nm(\lambda)\rangle$ and $E_{nm}(\lambda)$.
Finally, the $\lambda$-independent level populations $p_n$
now become $p_{nm}$.

For observables of the form (\ref{n10}) the time-averaged
expectation value (\ref{n70}) thus can be rewritten as
\begin{eqnarray}
\overline{\langle A \rangle (t)} & = & \sum_{mn} p_{nm}\,  _S\langle n|A^S|  n\rangle_S 
\nonumber
\\
& = &\sum_n p_n^S\, _S\langle n|A^S|  n\rangle_S
\label{705}
\\
p^S_n & := & \sum_m p_{nm} \ .
\label{704}
\end{eqnarray}

Next, we rewrite the $p_{nm}$ according to
(\ref{n85}) as
\begin{equation}
p_{nm}=h_\lambda (E_{nm}(\lambda))+\delta p_{nm} 
\label{n85a}
\end{equation}
Regarding the ``unbiased fluctuation'' $\delta p_{nm}$
(cf. Sect. \ref{s41}) can conclude that for 
any fixed index $n$, the sub-set $\delta p_{nm}$
with variable indices $m=0,1,2,...$
is unbiased as well, i.e.
\begin{eqnarray}
\sum_m p_{nm} = \sum_m h_\lambda(E_{nm}(\lambda)) \ .
\label{704b}
\end{eqnarray}
More precisely, the energies $E_{nm}$
from (\ref{12b}), with $n$ arbitrary but
fixed and $m$ variable, are still 
unimaginably dense, and the same
property is inherited by $E_{nm}(\lambda)$.
Further, there is no reason to expect the emergence
of any special ``correlations'' between the
ensemble averaged fluctuations
$\delta p_{nm}$ by selecting any 
sub-set with a fixed $n$ in (\ref{n85a}).

Note that during the preparation phase
(cf. Sect. \ref{s46a}), there 
may exists an indirect, non-weak coupling 
between system and bath in the canonical setup,
namely when both parts are interacting 
simultaneously with the rest of the world.
Then, it is even more obvious to expect that 
(\ref{704b}) will be canonically fulfilled.

Rewriting (\ref{n80}) as
$h_\lambda(E_{nm}(\lambda))=h_0(E_{nm})$
it follows with (\ref{704}) and (\ref{704b}) 
that $p_n^S=\sum h_0(E_{nm})$ and with
(\ref{n90}), (\ref{n100}) that $p_n^S=\sum \rho_\lambda(E_{nm})/\omega_0(E_{nm})$.
Dropping indices ``$0$'' corresponding to $\lambda=0$ as usual, 
we finally arrive at
\begin{equation}
p_n^S=\sum_m \frac{\rho(E_{nm})}{\omega(E_{nm})} \ .
\label{201a}
\end{equation}
where $\rho(E)$ stands for the ``true'' energy 
density $\rho_\lambda(E)$ of the 
system-plus-bath compound at finite coupling $\lambda$.

\subsubsection{Additivity of Entropy}
\label{s911}
Here, we revisit the issues of level counting,
entropy and temperatures from Sect. \ref{s21}
for the special Hamiltonian (\ref{11}) with 
$\lambda=0$.

The definitions and relations (\ref{1})-(\ref{8a}) 
can be taken over without any change
to our present special case, except that all
single indices $n$ now become double indices $nm$
(see (\ref{12a}), (\ref{12b})).
We reiterate that all those definitions and relations
remain for the moment on a purely formal level without
any reference to the actual system state.
Their only purpose at this stage is to count levels
in a convenient way.

Since $\lambda=0$ in (\ref{11}),
we are dealing with two individual isolated systems
and hence analogous definitions and equations as in
(\ref{1})-(\ref{8a}), but now with indices ``S'' and ``B'',
apply separately to the system and to the bath.
As detailed in Appendix A, the following relations 
between those separate system and bath quantities and the
original quantities for the system-plus-bath compound 
(\ref{11}) with $\lambda = 0$ can be established:
Focusing on $E>E_0$, we denote by $E_B(E)$ the
maximum of $S_B(E')+S_S(E-E')$ with respect to $E'$, i.e.
\begin{eqnarray}
E_B(E):=\max_{E'}\!\arg\{ S_B(E')+S_S(E-E')\} \ .
\label{a1}
\end{eqnarray}
In the generic case this maximum is unique and is
contained in the interval $(E_0,E)$.
Adopting the definition
\begin{eqnarray}
E_S(E):= E- E_B(E) \ ,
\label{a2}
\end{eqnarray}
the following results are established in Appendix A:
\begin{eqnarray}
S(E) & = & S_B(E_B(E))+S_S(E_S(E))
\label{23}
\\
T(E) & = & T_B(E_B(E))=T_S(E_S(E))
\label{24}
\end{eqnarray}
More precisely, these relations are asymptotically 
exact approximations for $f_B\to\infty$. 
But since $f_B$ is at least of the order of 
$10^{23}$ (see below (\ref{9a})) they are
satisfied with extremely high accuracy.

In other words, in the zero coupling limit
the entropies of the systems-plus-bath 
compound exhibit an additive behavior 
provided the total energy $E$ is distributed among system 
and bath such that $S_B(E')+S_S(E-E')$ is maximized, 
which in turn has the consequence that all 
temperatures are identical, i.e. the so-called
equilibrium condition (or zeroth law of thermodynamics) 
is fulfilled.

\section{Outlook on General non-interacting systems}
\label{s11}
Systems consisting of non-interacting particles 
or other types of non-interacting sub-systems
are popular models in many different contexts.
Also the canonical system-plus-bath setup from
the previous section is of this structure.
Strictly speaking, every single sub-system would 
thus be isolated and its energy would be a conserved 
quantity, thus prohibiting any kind of equilibration
or thermalization process between different sub-systems.
To each of them, the discussion of thermalization and
and the concomitant open questions from 
Sect. \ref{s8} apply. 
In particular, individual sub-systems which 
are ``small'' (e.g. single particles)
may not even exhibit equilibration
(cf. Sect. \ref{s1}).
Accordingly, the term ``non-interacting'' actually
means an interaction which is strictly speaking 
finite, thus giving rise to ``normal'' equilibration 
and thermalization, but in certain other respects 
still ``negligibly small''.
A more precise formulation of such a weak coupling 
condition and its implications analogously to 
Sect. \ref{s9} is straightforward.

Similarly as in Sect. \ref{s912} one finds that
the generalized non-resonance condition (\ref{52})
is violated at zero coupling \cite{lin09} but is generically 
restored as soon as the slightest interaction 
between the sub-systems is included.
Note that in Sect. \ref{s912} it is tacitly assumed 
that the two non-interacting sub-systems are 
distinguishable.
In the opposite case of indistinguishable 
sub-systems (e.g. indistinguishable particles),
all admitted states of the compound system
must be symmetric (Bosonic sub-systems)
or anti-symmetric (Fermionic sub-systems)
against exchanging the indices of the
two sub-systems, i.e. the pair $(n,m)$ 
has to be identified with (is indistinguishable from) 
$(m,n)$ for all $n,m$ in (\ref{12a}), (\ref{12b}).
Accordingly, in the identity
$E_{nn}-E_{nm}=E_{mn}-E_{mm}$
discussed below (88),
the two energies $E_{nm}$ and 
$E_{mn}$ must be identified.
Yet, condition (88)
is still violated.

These considerations demonstrate that the generalized
non-resonance condition
(\ref{52}) is a quite sensitive criterion
in the context of equilibration and thermalization.
It thus seem likely that a significantly weaker 
but still completely general condition of this 
type may not exist.

Note that a system of strictly non-interacting 
particles (or other kinds of sub-systems) 
which is coupled to a bath (canonical setup)
generically gives
rise to a total system-plus-bath compound
which cannot be decomposed into isolated
sub-systems any more, and therefore generically
fulfills condition (\ref{52}).

Beyond the realm of ``weak coupling'',
a partitioning of the total compound into 
physically meaningful sub-systems becomes 
questionable.
In particular, it does not make much sense 
to speak about properties of one 
``sub-system alone''.
Rather, the total compound amounts to an isolates 
system without any special properties, to which 
the general discussion form Sect \ref{s8} 
applies.

\section{Thermalization for the Canonical Setup}
\label{s7}
The objective of the present section is to 
establish thermalization without making 
use of the unproven property (\ref{p})
in the most important special case, namely
the canonical setup from Sect. \ref{s9}.

As detailed in Sect. \ref{s8},
it is taken for granted that the system energy 
\begin{eqnarray}
E^\ast:=\langle H\rangle
\label{104}
\end{eqnarray}
is known with high accuracy.
In other words, the energy density $\rho(E)$
from (\ref{710}) exhibits a very narrow peak within 
a close vicinity of $E^\ast$.
By combining (\ref{710}) and (\ref{104}) 
one recovers the expected relation
\begin{eqnarray}
E^\ast = \int dE\, E \, \rho(E) \ .
\label{717}
\end{eqnarray}
The ``star'' in $E^\ast$ emphasises the fact 
that the ensemble averaged energy 
of the ``real'' system is fixed.
Identifying $E$ in Sect. \ref{s911}
with $E^\ast$ establishes the connection
between the so far unrelated considerations in
Sect. \ref{s911} and our present issue of 
thermalization.

Formally, the task is to show that in (\ref{101})
the unknown details of $p_n$ are 
(practically) irrelevant. 
In particular, this then implies
that (\ref{101}) indeed agrees with the prediction
of equilibrium Statistical Mechanics.

As detailed in Sect. \ref{s9}, the canonical setup 
consists of a ``small'' system (S) which is 
weakly coupled to a much ``bigger'' bath (B).
Equilibration of the isolated system-plus-bath 
compound is taken for granted, i.e. 
expectation values (32) become practically 
indistinguishable from (\ref{101}).
Focusing on system observables of the form
(\ref{n10}) and observing (\ref{n20}) and (\ref{705}),
those equilibrium expectation values thus take the form
\begin{eqnarray}
\langle A\rangle & = & \tr_S\{\rho_{eq}^S\, A^S\} 
\label{705a}
\\
\rho_{eq}^S & = &  \sum_n p_n^S\ |n\rangle_{S\, S}\langle n|
\label{703}
\end{eqnarray}
where $\tr_S$ indicates the trace in $\hr_S$.
In other words, as far as system properties 
are concerned, the knowledge of the reduced
equilibrium density operator $\rho_{eq}^S \, :\, \hr_S\to\hr_S$ 
is sufficient.

\subsection{Boltzmann-form of $p_n^S$ and canonical density}
\label{s71}
Exploiting (\ref{201a}) 
we can conclude that
\begin{eqnarray}
p^S_n  =  \int dE\ \frac{\rho(E)}{\omega(E)}\ \sum_m \delta(E-E_{nm})
\label{201}
\end{eqnarray}
where the delta-function is, as usual, 
considered as washed out.
The first term under the integral can be
rewritten by means of (\ref{6}) 
\begin{eqnarray}
\frac{\rho(E)}{\omega(E)}=\rho(E)\frac{k_B T(E)}{\Omega(E)} \ .
\label{202}
\end{eqnarray}
Turning to the second term under the integral,
we exploit (\ref{12b}), the definition of 
$\omega_B(E)$ analogous to (\ref{4}), and the
corresponding relation (\ref{6}), yielding
\begin{eqnarray}
& & \sum_m \delta(E-E_{nm}) = \sum_m \delta(E-E^S_{n}-E^B_{m}) =
\nonumber
\\ 
& & = \omega_B(E-E^S_{n})=\frac{\Omega_B(E-E^S_{n})}{k_B T_B(E-E^S_{n})}
\label{203}
\end{eqnarray}
Making use of (\ref{2}), (\ref{202}), and (\ref{203}) we
can rewrite (\ref{201}) as
\begin{eqnarray}
p^S_n & = &  \int dE\ \rho(E)\ \frac{T(E)}{T_B(E-E^S_{n})}\ e^{Q(E)}
\label{204}
\\
Q(E) & := & \frac{S_B(E-E_n^S)-S(E)}{k_B}
\label{205}
\end{eqnarray}
where the dependence of $Q$ on $n$ has been dropped.
Exploiting (\ref{a2}) and (\ref{23}) we see that
\begin{eqnarray}
k_B\, Q(E) & = & S_B(E_B(E)+E_S(E)-E_n^S)
\nonumber
\\
& & -S_B(E_B(E))-S_S(E_S(E)) \ .
\label{206}
\end{eqnarray}
According to the mean value theorem there exists 
for any given $x$-and $y$-value a $\vartheta\in[0,1]$
with the property that
\begin{eqnarray}
S_B(x + y) = S_B(x) + y\ S_B'(x + \vartheta y)
\label{901}
\end{eqnarray}
Choosing $x=E_B(E)$ and $y=E_S(E)-E_n^S$ and exploiting
(\ref{5}), we can rewrite (\ref{206}) as
\begin{eqnarray}
Q(E) & = & \frac{E_S(E) - E^S_n}{k_B T_B(E_B(E)+\Delta)} - \frac{S_S(E_S(E))}{k_B}
\label{902}
\\
\Delta & := & \vartheta\, [E_S(E)-E_n^S] \ ,
\label{903}
\end{eqnarray}
where the dependence of $\vartheta$ and $\Delta$
on $E$ and $n$ has been dropped.

We first consider the simplest case of a delta-distributed
energy density
\begin{equation}
\rho (E)=\delta(E-E^\ast) \ .
\label{903a}
\end{equation}
Thus, (\ref{204}) takes the form
\begin{eqnarray}
p^S_n & = & \frac{T(E^\ast)}{T_B(E^\ast-E^S_{n})}\ e^{Q(E^\ast)}
\label{903b}
\end{eqnarray}
Observing (\ref{a2}), the denominator $T_B(E^\ast-E^S_{n})$ 
can first be rewritten as $T_B(E_B(E^\ast)+E_S(E^\ast)-E^S_{n})$
and then with (\ref{8}) as
\begin{equation}
T_B(E^\ast-E^S_{n})=T_B(E_B(E^\ast))
\, \left[1+\ord\left(\frac{E_S(E^\ast)-E^S_{n}}{E_B(E^\ast)-E^B_0}\right)\right] 
\label{903c}
\end{equation}
Finally, with (\ref{24}) and relations like in (\ref{7}) but
with indices S and B we can conclude that
\begin{equation}
T_B(E^\ast-E^S_{n})=T(E^\ast)
\, \left[1+\ord\left(\frac{f_S}{f_B}\frac{E_S(E^\ast)-E^S_{n}}{E_S(E^\ast)-E^S_0}\right)\right] 
\label{903d}
\end{equation}
In view of (\ref{9a}), the last summand is negligible and (\ref{903b})
takes the form
\begin{eqnarray}
p^S_n & = & e^{Q(E^\ast)}
\label{903e}
\end{eqnarray}
Similarly as in (\ref{903d}), one sees that 
$T_B(E_B(E)+\Delta)$ appearing in (\ref{902}) 
can be approximated by $T_B(E_B(E))=T(E)$,
yielding
\begin{eqnarray}
Q(E^\ast) & = & \frac{E_S(E^\ast) - E^S_n}{k_B T(E^\ast)} 
- \frac{S_S(E^\ast_S(E))}{k_B} \ .
\label{903f}
\end{eqnarray}
With the usual definitions of the free energy $F_S(E)$ and the partition
sum $Z_S(E)$ of the system $S$, namely
\begin{eqnarray}
F_S(E) & := & E_S(E)-T_S(E)\, S_S(E)
\label{903g}
\\
Z_S(E) & := & e^{-F_S(E)/k_B T_S(E)}
\label{903h}
\end{eqnarray}
in combination with (116), we can rewrite (115) as
\begin{eqnarray}
p^S_n & = & \frac{1}{Z_S(E^\ast)}\  e^{-E_n^S/k_BT(E^\ast)} \ .
\label{903i}
\end{eqnarray}
Taking into account the normalization condition
$\sum_n p_n^S=1$, we recover 
\begin{eqnarray}
Z_S(E^\ast) = \sum_n e^{-E_n^S/k_B T(E^\ast)} \ .
\label{903j}
\end{eqnarray}

We finally turn to general energy densities $\rho(E)$.
Since $\rho(E)$ enters linearly in (\ref{204}),
we simply can superimpose the results (\ref{903i})
for sufficiently many delta-functions 
approximating the true $\rho(E)$.
The fact that each delta-function brings along
a somewhat different value of $E^\ast$ in (\ref{903i})
has a negligible effect according to (\ref{8})
as long as $\rho(E)$ is still sharply peaked
about its mean value.
Denoting this mean value, in accordance with
(\ref{104}) and (\ref{717}), again by the 
symbol $E^\ast$, 
one thus recovers exactly the same relations as
in (\ref{903g})-(\ref{903j}).

In summary, the canonical formalism 
(\ref{903g})-(\ref{903j}) is valid 
in full generality.
Apart from the average energy $E^\ast$,
all the remaining details of the
(unknown) energy density $\rho(E)$ 
do not matter.
The Boltzmann-distribution (\ref{903i})
together with (\ref{703}) yields the
canonical density operator
\begin{equation}
\rho_{eq}^S = \frac{1}{Z_S(E^\ast)}\ e^{-H_S/k_B T(E^\ast)} \  ,
\label{906}
\end{equation}
where $T(E^\ast)$ is the temperature corresponding
to the given total energy $E^\ast$ of the isolated
system-plus-bath compound.
In practice, this energy $E^\ast$  
is usually not known, and one thus rather considers
the temperature $T$ as ``given''.
Accordingly, in (\ref{903i})-(\ref{906}) the state function
$T(E^\ast)$ is replaced by the ``new'' independent state
variable $T$ and similarly $Z_S(E^\ast)$ by $Z(T):=Z_S(E^\ast(T))$.

\section{Summary and Conclusions}
\label{s10}
In the first part of this paper we considered 
general, isolated quantum
systems with many degrees of freedom
$f$, and being extensive in the 
sense of Eqs. (\ref{3}) and (\ref{7}).
As a further ``generic'' property of the
Hamiltonian $H$, the (generalized) non-resonance condition
(\ref{52}) was taken for granted.
Our key assumptions concerning the
``realistic modeling'' of actual experimental
systems were: (i) observables have a 
``reasonably bound'' range-to-resolution ratio and
(ii) initial conditions may be arbitrarily 
out of equilibrium but, on the
average over the entire statistical ensemble
(many repetitions of the ``same'' experiment),
they give rise to a  well-defined
population density
(average occupation probability of 
many neighboring energy levels).
The latter assumption seems quite plausible 
{\em per se}, but can also be justified 
via the experimental preparation procedure 
at the origin of the initial condition.

All further ``details'' of the initial condition 
and the Hamiltonian were left unspecified, 
reflecting the unavoidable actual lack 
of knowledge in this respect.

Given the initial condition, the exact
standard Quantum Mechanical time evolution 
was adopted without any approximation or 
modification.

Our first main result (\ref{77}) implies that
after initial transients have died out, 
the system looks for all practical purposes 
{\em as if} it were in a steady state described 
by the so-called generalized Gibbs ensemble 
$\rho_{eq}$, in spite 
of the fact that the ``true'' density operator
$\rho(t)$ never becomes stationary, but rather 
exhibits the well-known Quantum Mechanical 
recurrence and time inversion invariance 
properties.
Our key conclusion was that the mathematically 
undeniable differences between
the ``apparent equilibrium'' $\rho_{eq}$ 
and the ``true'' density operator $\rho(t)$
are either unobservably small or 
unobservably rare in time.

While the issue of equilibration can thus be
considered as settled, that of thermalization
still remains an open problem as far as completely
general isolated systems are concerned, as
detailed in Sect. \ref{s8}.

In the second part of the paper we focused on
a special case of foremost practical relevance,
namely the canonical setup, consisting of
a system of actual interest (that may be 
macroscopic or not) which is weakly coupled
to a much ``bigger'' environment.
Provided the total system-plus-bath compound
satisfies the above conditions for 
equilibration, the corresponding ``apparent
equilibrium'' $\rho_{eq}$ reduces, after
eliminating (tracing out) the bath, to the
canonical density operator (\ref{906}), 
independently of all the unknown 
``microscopic details'' of the possibly 
far from equilibrium initial condition.
In other words, the ``small'' system 
is proven to exhibit ``thermalization''.
The result even goes beyond the claim
of Statistical Mechanics in so far as
not only the system but also the bath
may be initially out of equilibrium.

It seems not unlikely that our main prerequisites
in deriving these results cannot be 
substantially weakened any more:
Systems which can be decomposed into strictly
non-interacting sub-units 
(e.g. non-interacting particles)
are known not to thermalize, 
and indeed violate the non-resonance 
condition (\ref{52}).
Likewise, when either an unlimited
range-to-resolution ratio or an 
initial condition without a well-defined
population density is admitted, one
readily finds examples which do
not exhibit equilibration, see 
Sects. \ref{s2a} and \ref{s47}.

While our main focus has been on statistical
ensembles, we have noticed in Sect. \ref{s63}
that all the above results also remain
valid for pure states $|\psi (t)\rangle$,
provided the initial energy level
occupation probabilities
$|\langle \psi(0)|n\rangle|^2$ satisfy the 
condition that a well-defined
population density exists.
Specifically, within the canonical setup
the reduced system density operator
will again be practically indistinguishable
from the canonical ensemble for the 
overwhelming majority of times $t$.
This result is closely related to the
issue of ``canonical typicality''
from Refs. 
\cite{pop06,rei07,rei07a,gol06,gol06a,bar09}.

Next, we briefly address the issue of low
temperatures.
First, for extremely low temperatures, the
rough estimates from (\ref{3}) and (\ref{7})
may break down. Since these estimates are at 
the heart of our present approach, also our 
main results may not be valid any more.
Essentially this happens when the
entropy becomes experimentally
indistinguishable from zero.
This may, but need not be the case 
for Bose-Einstein condensates \cite{exp}.

Further, the common notion
that a Bose-Einstein condensate
exhibits a macroscopically populated 
ground state may be easily misunderstood
in our present context.
Namely, this notion refers to the 
fact that the total many particle
product state contains a large 
number of single particle ground 
states.
The word ground state thus refers 
to the individual (non-interacting) 
particles, not to the total many
particle system.
Indeed, besides the numerous single particles 
in their individual ground state, 
there may still remain many
further particles which are in 
excited single particle states.
Hence, we are in fact not dealing with the
actual ground state of the many particle
product Hilbert space.
Rather, it may easily happen that
the maximal population of all the
many particle product states is 
still small and hence the conditions
regarding the level populations from 
Sect. \ref{s4} may still be satisfied.

We close with a few remarks regarding 
the issues of (non-)integrability, 
ergodicity, chaos, 
decoherence, and entanglement.
We first remark that (non-)integrability,
ergodicity, and chaos are relatively
well defined notions for classical systems,
but that their role with respect to 
equilibration and thermalization is not
really clear in the classical case.
The corresponding notions in the realm of 
quantum systems are much less well and
uniquely defined \cite{rig08,wei92}.
But even if this problem would be solved, 
in view of the classical situation,
the usefulness of those concepts for 
equilibration and thermalization 
are likely to be limited
in the quantum case as well.

The closest connection of our present 
approach to the above notions may be via 
the non-resonance condition (\ref{52}).
However, such a connection or analogy
does not seem to offer any additional 
physical insight.
Concerning our requirement that the
initial condition must exhibit a well
defined population density,
we remark that during the 
preparation phase (which represents
the physical origin of the initial 
condition), a distinction between
integrable and non-integrable 
systems does not make much 
sense anyhow.

The concepts of integrability, ergodicity
and the like may well play a crucial role
for the following two issues:
(i) The transient relaxation-process of 
the initial state towards equilibrium,
both qualitatively (exponential decay or 
not) and quantitatively (estimating
the relaxation time).
This is suggested by the well-established
role of level statistic in quantum chaos
and the importance of energy differences
throughout the present Appendix D.
(ii) The general problem of thermalization
addressed in Sect. \ref{s8},
in particular the unproven key postulate 
(\ref{p}) in this context.

As far as the issue of equilibration is 
concerned, our present approach and results
demonstrate that entanglement and decoherence
play no role since we are dealing with 
isolated systems without any external 
influence by the rest of the world.
With respect to thermalization, the
question remains open.

\subsection{Comparison with related works}
\label{s101}
We first address several works with a finite
but not too small overlap with the present
one and then turn to the two most
closely related Refs. \cite{lin09,rei08}.
The pertinent literature regarding 
the ``missing link'' (\ref{p})
in the context of thermalization
has already been addressed 
at the end of Sect. \ref{s8}.

Considering and estimating quantities like
(\ref{65}) is very natural and has a long 
tradition:
Merits and shortcomings of the early 
works are reviewed e.g. in \cite{ergod},
most notably Ludwig's approach \cite{lud58}.
In particular, many of them \cite{boc59,ergod}
involve an extra average over initial
conditions with the effect that any 
specific non-equilibrium initial condition
(representing a given experiment) 
must be excluded as ``potentially untypical'' 
from the general conclusions.
Turning to the more recent precursors,
Peres' approach \cite{per84} is roughly
comparable to ours up to Eq. (\ref{69}) but 
then proceeds with the conjecture 
that the $\tilde A_{mn}$ are pseudorandom 
matrix elements, statistically 
independent of the $\rho_{nm}$,
for which there are general arguments 
\cite{per84} and numerical evidence 
\cite{fei84} 
(and counter-evidence \cite{rig08})
but no proof.
For {\em pure} states, Srednicki obtained 
similar results \cite{sre96} by exploiting 
a common conjecture about the semiclassical 
behavior of classically smooth observables 
$A$ in systems with a fully chaotic classical 
limit.
Again, this conjecture is based on good 
arguments \cite{arg} but no proof.
Moreover, typical classical many-body systems 
are not expected to behave fully chaotic
\cite{bun08,skl93}.
Somewhat similar conclusion have been reached
even earlier by Deutsch \cite{deu91} via 
additional hand waving arguments.
Finally, rigorous results comparable
to (\ref{76}) are due to \cite{tas98,cra08},
but only for rather special Hamiltonians 
$H$ and initial conditions.
Within the same restrictions, the
ground breaking work by Tasaki \cite{tas98} 
also addresses the issue of thermalization 
by arguments which are similar in spirit 
to those in our present Sect. \ref{s9}.

The first part of the present work
(until the end of Sect. \ref{s8})
represents a generalization and more 
detailed explanation of the Letter 
\cite{rei08}.
The main extensions consist in the
enlarged class of observables admitted
in Sect. \ref{s3} and the fact that
degenerate energy eigenvalues are not
any more excluded in our present 
work.

We finally turn to the closely related 
work \cite{lin09}.
In contrast to our present work,
Ref. \cite{lin09} is focused on
Hilbert spaces which are finite 
dimensional and which exhibit a 
``system-plus-bath'' product
structure of the form (\ref{10}).
Apart from a non-resonance condition 
(excluding degeneracies, see below 
(\ref{52})), the Hamiltonian 
may still be completely arbitrary.
Further, the system is assumed to be in a 
{\em pure state} on the total system-plus-bath 
Hilbert space,
while the obtained results mainly
concern the reduced (usually mixed) 
state of the ``small'' system
after tracing out the ``large'' bath.
Apart from these quite significant
overall differences, the main findings
with respect to equilibration are
rather similar in character to ours.
In particular, observables with finite 
range are implicitly taken for granted
according to the discussion below 
Eq. (3) in Ref. \cite{lin09}, and the
``effective dimension'' $d_{eff}$ from
\cite{lin09} is basically equivalent to
$\tr\{\rho_{eq}^2\}$ in our present 
approach, as discussed below (\ref{73}).
With respect to thermalization,
the results from \cite{lin09} are
of a quite different character than 
ours, mostly concerning ``typical'' 
\cite{pop06,gol06,bar09}
properties of pure states which are 
randomly sampled according to a uniform 
probability density from 
certain sub-Hilbert spaces.
In the opinion of the present author
(see also Sects. \ref{s1} and \ref{s46a}),
the main open question of this approach 
is in how far one particular pure state 
or an ensemble of uniformly distributed 
pure states are suitable to
describe a real experimental setup.

\subsection*{Acknowledgment}
Special thanks is due to Michael Kastner and Roderich 
Tumulka for pointing out Refs. \cite{lof,geo95}
and to an anonymous referee for insisting in a more
detailed discussion of the notion of weak coupling 
(Sect. \ref{s9}).

\section{Appendix A}
\label{app1}
In this Appendix we establish the relations
(\ref{23}) and (\ref{24}).
As detailed above (\ref{a1}), analogous definitions and 
equations as in (\ref{1})-(\ref{8a}), but with indices 
``S'' and ``B'', hold true and are exploited 
in the following.

With the help of (\ref{4}) and (\ref{12b}) we can conclude that
\begin{eqnarray}
\!\!\!\!\!\!\!\!\! \omega(E) & = & \sum_{nm}\delta(E-E_n^s-E_m^B)=\sum_n \omega_B(E-E_n^S)=
\nonumber
\\
\!\!\!\!\!\!\!\!\! & = & \int dE' \, \omega_B(E-E') \, \sum_n \delta(E'-E_n^S) =
\nonumber
\\
\!\!\!\!\!\!\!\!\! & = & \int dE' \, \omega_B(E-E') \, \omega_S(E') \ ,
\label{13}
\end{eqnarray}
where $\sum_{nm}$ indicates a summation over all $n,m=0,1,2,...$.
It follows from (\ref{4}) that $\omega_S(E')=0$ for $E'<E_0^S$ and 
similarly that $\omega_B(E-E')=0$ is for $E'>E-E_0^B$.
Taking for granted that the integrand $\omega_B(E-E') \, \omega_S(E')$ 
is a sufficiently smooth function of $E'$, it thus will have an
absolute maximum in the interior of the interval $[E_0^S,\, E-E_0^B]$.
For any given $E$-value exceeding the ground state energy
$E_{00}=E_0^S+E_0^B$ of the compound system, the absolute
maximum will furthermore be generically unique.

Next we rewrite (\ref{13}) by means of (\ref{3}) and (\ref{6}) as
\begin{eqnarray}
\!\!\!\!\!\!\!\!\! 
\frac{e^{S(E)/k_B}}{k_B T (E)}
 & = &  \int dE' \, 
\frac{e^{[S_B(E')+S_S(E-E')]/k_B}}{k_B T_B(E')k_BT_S(E-E')}
\label{14}
\end{eqnarray}
Similarly to the integrand in (\ref{13}), the exponent in (\ref{14})
generically exhibits a unique absolute maximum
at some $E'$-value in the interior of $[E_0^S,\, E-E_0^B]$,
henceforth denoted as $E_B(E)$.
The next step is to evaluate (\ref{14}) by means of a saddle
point approximation, i.e. by expanding the exponent around its
maximum up to the second order.
At the maximum, the derivative of the exponent vanishes,
yielding with (\ref{5}) the relation
\begin{equation}
T_B(E_B(E))=T_S(E_S(E))
\label{15}
\end{equation}
where we have introduced
\begin{equation}
E_S(E):=E-E_B(E)
\label{16}
\end{equation}

We emphasise that, like in Sects.
\ref{s21} and \ref{s911}, 
we avoid speaking about 
system states.
If we give up this viewpoint for a second, then
(\ref{15}) is nothing else than the 
so-called equilibrium condition for two systems
with negligible (but non-zero) interaction
in thermal equilibrium, sharing the total
system energy $E$ according to (\ref{16}).

Expanding the exponent on the right hand side of (\ref{14})
up to the second order about its maximum at $E'=E_B(E)$
in combination with (\ref{8a}) yields
\begin{eqnarray}
& & 
\frac{S_B(E')+S_S(E-E')}{k_B}
=\frac{S_B(E_B(E))+S_S(E_S(E))}{k_B}
\nonumber
\\
& - & \ord\left(f_B\left[\frac{E'-E_B(E)}{E_B(E)-E^B_0}\right]^2
+
f_S\left[\frac{E'-E_B(E)}{E_S(E)-E^S_0}\right]^2\right)
\label{17}
\end{eqnarray}
Since $f_B$ is at least of the order of $10^{23}$
(see below (\ref{9a})), it follows that only
an extremely small neighborhood of the maximum
notably contributes to the integral in (\ref{14}),
and within this neighborhood the variations of
the non-exponential factors on the right hand side of 
(\ref{14}) are
negligibly small according to (\ref{8}).
Performing the remaining Gaussian integral
in (\ref{14}) yields
\begin{eqnarray}
\frac{S(E)}{k_B} & = & \frac{S_B(E_B(E))+S_S(E_S(E))}{k_B}+ \ln(\ord(R))
\label{20}
\\
R & = & \sqrt{r}\, k_BT(E)/[k_B T_B(E_B(E))]^2
\label{21}
\\
r & = & \frac{f_B}{[E_B(E)-E^B_0]^2}+\frac{f_S}{[E_S(E)-E^S_0]^2}
\label{22}
\end{eqnarray}
From (\ref{7}) and (\ref{9}) one can infer that $R$ is of the
order of $f$.
Considering that the quantity in (\ref{20}) scales like
$f$ according to (\ref{7}) and that $f$ is at least of the 
order of $10^{23}$, the contribution of $\ln(\ord(R))$ 
in (\ref{20}) is negligible.
We thus obtain in extremely good approximation the
relation
\begin{eqnarray}
S(E) & = & S_B(E_B(E))+S_S(E_S(E))
\label{23a}
\end{eqnarray}
Differentiating this relation with respect to $E$ and taking into
account (\ref{5}), (\ref{15}), and (\ref{16}) yields
\begin{equation}
T(E)=T_B(E_B(E))=T_S(E_S(E))
\label{24b}
\end{equation}
The latter two relations are identical to
(\ref{23}) and (\ref{24}) in the main text

\section{Appendix B}
\label{app2}
In this Appendix we derive the relation
\begin{equation}
\sum_n \rho_{nn}^2(0) \leq \max_{n}p_{E_n} \ .
\label{281} 
\end{equation}
To do so, we first exploit (\ref{278}) to conclude
\begin{eqnarray}
p_{E_n}^2 & = &  \sum_{E_m=E_n}\sum_{E_m'=E_n} \rho_{mm}(0) \rho_{m'm'}(0)
\nonumber
\\
& \geq & \sum_{E_m=E_n} \rho_{mm}^2(0)
\label{279} 
\end{eqnarray}
and hence
\begin{eqnarray}
\sum_n \rho_{nn}^2(0) & = & 
\sum_{E_n} \sum_{E_m=E_n} \rho_{mm}^2(0)
\nonumber
\\
& \leq & \sum_{E_n} p_{E_n}^2 \leq \sum_{E_n} p_{E_n}\max_{m}p_{E_m} \ .
\label{280} 
\end{eqnarray}
With (\ref{275}) we obtain (\ref{281}).

\section{Appendix C}
\label{app3}
In this Appendix, the derivation of
(\ref{716}) is provided.
To start with, we recall that the delta-functions
in (\ref{710}) and (\ref{712}) are understood
to be ``washed out'' over many energy levels.
Hence, smoothening $\rho(E)$ with the help of 
yet another washed out delta-function actually 
does not change $\rho(E)$ any more:
\begin{eqnarray}
\rho(E)=\int dE'\ \delta(E-E')\ \rho(E) \ .
\label{713}
\end{eqnarray}
Introducing (\ref{706}) and (\ref{712}) on the right
hand side yields
\begin{eqnarray}
\rho(E) & = & \int dE'\ \delta(E-E')\sum_{n}h(E_{n}) \delta(E'-E_{n})
\nonumber
\\ 
& + & \sum_{n}\delta p_{n} \delta(E-E_{n}) \ .
\label{714}
\end{eqnarray}
Since the delta-functions are washed out, the
last summand essentially amounts to a local average over
many $\delta p_{n}$ and is thus negligible (see below (\ref{706})).
In turn, the function $h(E)$ hardly changes within 
the peak region of the delta-functions. Hence $h(E_{n})$
can be replaced by $h(E')$ and then by
$h(E)$. Altogether we thus obtain
\begin{eqnarray}
\rho(E) & = & h(E) \int dE'\ \delta(E-E')\sum_{n}\delta(E'-E_{n}) \ .
\label{715}
\end{eqnarray}
The sum can be identified with $\omega(E')$ from (\ref{4}).
Since the latter is once again already a locally averaged quantity,
the integral over the last remaining 
delta-function is trivial, yielding (\ref{716}).

\section{Appendix D}
\label{app4}
According to (\ref{62}) it follows that
\begin{eqnarray}
\overline{e^{iat}}  & = &  1 \ \mbox{for $a=0$}
\nonumber
\\
\overline{e^{iat}}  & = &  0 \ \mbox{for $a\not=0$} \ .
\label{64x}
\end{eqnarray}
Next we exploit (\ref{62}) and (\ref{27}) to conclude that
\begin{equation}
\overline{\rho(t)}=\sum_{mn} \rho_{mn}(0)\, \overline{e^{-i[E_m-E_n]t/\hbar}} 
\, |m\rangle\langle n| \ .
\label{64}
\end{equation}
In combination with (\ref{277}) and (\ref{64x}) we thus 
recover (\ref{61}).

With (\ref{28}) and (\ref{64b}) we can rewrite
the variance from (\ref{65}) as
\begin{eqnarray}
\sa^2 & = & \overline{ [\tr\{\rho (t)A \}-\tr\{\rho_{eq} A\}]^2} 
\nonumber
\\
& = & \overline{ [\tr\{\tilde \rho(t)A\} ]^2}
\label{66}
\\
\tilde\rho(t) & := & \rho(t)-\rho_{eq} \ .
\label{66a}
\end{eqnarray}
Exploiting (\ref{27}), (\ref{277}), and (\ref{61}) we obtain
\begin{equation}
\tilde\rho(t)={\sum_{mn}}' \rho_{mn}\, e^{-i[E_m-E_n]t/\hbar}
\, |m\rangle\langle n| \ ,
\label{66e}
\end{equation}
where ${\sum_{mn}}'$ indicates a summation over all 
$m,n=0,1,2...$ with $E_m\not =E_n$,
and where we adopted the abbreviation
\begin{eqnarray}
\rho_{mn}:= \rho_{mn}(0) \ .
\label{67b}
\end{eqnarray}

It follows that $\tr \{ \tilde\rho(t)\} = 0$ and hence
\begin{equation}
\tr\{\tilde \rho(t)(A + 1_\hr\, c)\} = \tr\{\tilde \rho(t)A\} 
\label{66d}
\end{equation}
for any $c\in\RR$, where
$1_\hr$ is the identity on $\hr$.

As said below (\ref{28a}), we can and will
replace $\hr$ and $A$ by $\hr_+$ and $A_+$ 
in the rest of this Appendix. 
Introducing 
\begin{equation}
\tilde A := A_+ - 1_{\hr_+}\min_{\hr_+}\langle\psi|A|\psi\rangle
\label{66b}
\end{equation}
we can infer from (\ref{31a}) that 
\begin{equation}
0\leq \langle \psi|\tilde A|\psi\rangle \leq \da'\
\mbox{for all normalized}\  |\psi\rangle \in\hr_+ \ .
\label{67}
\end{equation} 
Taking into account (\ref{66d}) and (\ref{66b}),
the variance (\ref{66}) can be rewritten as 
\begin{eqnarray}
\sa^2 & = & \overline{[\tr\{\tilde\rho(t)\tilde A\} ]^2} \ .
\label{67a}
\end{eqnarray}
Introducing (\ref{66e}) into (\ref{67a})
we obtain
\begin{equation}
\sa^2={\sum}' \tilde A_{jk} \rho_{kj} \tilde A_{mn} \rho_{nm} 
\, \overline{e^{i[E_j-E_k+E_m-E_n]t/\hbar}} \ ,
\label{68}
\end{equation}
where ${\sum_{jkmn}}'$ indicates a summation over all 
$j,k,m,n=0,1,2...$ with $E_j\not =E_k$ and $E_m\not =E_n$.

Since, according to (\ref{64x}), 
the time averaged exponentials in (\ref{68}) 
vanish if $E_j-E_k+E_m-E_n\not = 0$ we can conclude 
from the non-resonance condition (\ref{52}) that
\begin{eqnarray}
\sa^2={\sum_{mn}}' |\tilde A_{mn}|^2 |\rho_{mn}|^2 \leq
\sum_{mn} |\tilde A_{mn}|^2 |\rho_{mn}|^2  \ ,
\label{69}
\end{eqnarray}
where the first sum runs over all $m,n$ with 
$E_m \not=E_n$ and the second over all $m,n$. 
With (\ref{27a}) and (\ref{61}) we thus obtain
\begin{eqnarray}
\sa^2 & \leq & \sum_{mn} \tilde A_{mn} \rho_{nn} \tilde A_{nm} \rho_{mm}  
\nonumber
\\
& = &
\sum_{mn}\langle m|\tilde A\rho_{eq}|n\rangle \langle n|\tilde A\rho_{eq}|m\rangle  \ .
\label{70}
\end{eqnarray}
The sum over $n$ amounts to an identity operator and
that over $m$ yields 
\begin{eqnarray}
\sa^2 & = 
\tr\{[\tilde A\rho_{eq}]^2\} \ .
\label{70a}
\end{eqnarray}
Next, we evaluate this trace with the help of the eigenvectors 
$|\chi_n\rangle$ of $\tilde A$, yielding
\begin{eqnarray}
\sa^2\leq \sum_{mn} \langle \chi_m|\rho_{eq}\tilde A|\chi_n\rangle 
\langle \chi_n|\rho_{eq}\tilde A|\chi_m\rangle \ .
\label{71}
\end{eqnarray}
Observing that 
$\tilde A|\chi_n\rangle = |\chi_n\rangle\langle\chi_n| \tilde A|\chi_n\rangle$ 
(since $|\chi_n\rangle$ is eigenvector of $\tilde A$)
we can exploit (\ref{67}) to obtain
\begin{eqnarray}
\sa^2\leq (\da') ^2 \sum \langle \chi_m|\rho_{eq}|\chi_n\rangle 
\langle \chi_n|\rho_{eq}|\chi_m\rangle \ .
\label{72}
\end{eqnarray}
The sum over $n$ yields the identity operator
and that over $m$ amounts to $\tr\{\rho_{eq}^2\}$, yielding
\begin{eqnarray}
\sa^2 \leq (\da')^2\, \tr\{\rho_{eq}^2\} \ .
\label{73a}
\end{eqnarray}

Finally, we note that according to
(\ref{61c}) and (\ref{66a})
we can subtract from
$A$ in (\ref{66}) an arbitrary function
$B({\bf b})$ of the form (\ref{61b})
with the only consequence
in the final result (\ref{73a}) that $\da'$ 
goes over into $\daf'$.
Since this conclusion holds for arbitrary
$B({\bf b})$, the inequality even remains true
after minimization over all $B({\bf b})$,
i.e. we can replace $\da'$ in (\ref{73a})
by $\da''$ from (\ref{31b}).
In other words, we recover (\ref{73}).

\section{Appendix E}
\label{app5}
The purpose of this Appendix is a more detailed
justification of the non-degeneracy assumption 
adopted in Eq. (\ref{n50}).
More precisely, we will argue that the 
energy levels $E_n(\lambda)$,
considered as functions of $\lambda$, do not cross
each other for generic Hamiltonians $H(\lambda)$.

Intuitively, one expects that
the $E_n(\lambda)$ will be highly non-trivial functions of $\lambda$ and 
that for different $n$ these functions will behave notably ``different'' 
from each other.
Since the energy levels are unimaginably dense (cf. Sect. \ref{s21}),
crossings of neighboring levels $E_n(\lambda)$ upon variation of 
$\lambda$ thus might seem to be almost unavoidable.

But closer inspection shows that, on the contrary, such level crossings 
are actually avoided for generic Hamiltonians $H(\lambda)$ due to the
so-called level-repulsion mechanism:
The levels $E_n(\lambda)$, considered as functions of $\lambda$,
are governed by the following exact evolution equation, originally due to
Pechukas and Yukawa \cite{pec83}:
\begin{eqnarray}
\frac{d^2}{d\lambda^2}E_n(\lambda) 
& = & 2
\sum_{m\not=n} \frac{|V_{nm}(\lambda)|^2}{E_n(\lambda)-E_m(\lambda)}
\label{p1}
\\
V_{nm}(\lambda) 
& := & 
\langle n(\lambda) | \frac{d}{d\lambda} H(\lambda) | m(\lambda)\rangle \ ,
\label{p2}
\end{eqnarray}
where $E_n(\lambda)$ and $|n(\lambda)\rangle$ are the ``accompanying''
eigenvalues and eigenvectors of $H(\lambda)$.
The closing evolution equations for $V_{nm}(\lambda)$ are also known,
but not explicitly needed for our purpose.
The main point is that looking upon $\lambda$ as ``time'' and
$E_n(\lambda)$ as ``particle positions'', Eq. (\ref{p1}) is nothing else
than the Newtonian equation of motion for a one-dimensional ``gas''
of point particles, the so-called Pechukas-Yukawa gas \cite{pec83}.
The particles are repelling each other with ``coupling strengths''
$|V_{nm}(\lambda)|^2$ which depend on ``time'' $\lambda$.
In the generic case
(no special symmetries or ``selection rules''), the coupling
of two neighboring particles, say 
$|V_{nn+1}(\lambda)|^2$, will be positive with the exception of at most
a discrete set of time-points $\lambda$, and as a consequence,
any ``attempt'' of the two neighboring
levels $E_n(\lambda)$ and $E_{n+1}(\lambda)$ to cross each other is inhibited
by a repulsive ``force'' term in (\ref{p1})
of the form $|V_{nn+1}(\lambda)|^2[E_n(\lambda)-E_{n+1}(\lambda)]^{-1}$
which diverges as $E_n(\lambda)\to E_{n+1}(\lambda)$.



\begin{thebibliography}{7}


\bibitem{fey}
R. P. Feynman, Statistical Mechanics, 
Benjamin, Reading, Mass. 1972

\bibitem{rig08}
M. Rigol, V. Dunjko, and M. Olshanii, Nature {\bf 452}, 854 (2008)

\bibitem{lud58}
G. Ludwig, Z. Phys. {\bf 150}, 346 (1958); {\bf 152}, 98 (1958)

\bibitem{boc59}
P. Bocchieri and A. Loinger, Phys. Rev. {\bf 114}, 948 (1959)

\bibitem{ergod}
R. Jancel, {\em Foundations of Classical and Quantum Statistical Mechanics},
Pergamon, London (1969); 
I. E. Farquhar, {\em Ergodic Theory in Statistical Mechanics},
Interscience, NY (1964)

\bibitem{per84}
A. Peres, Phys. Rev. A {\bf 30}, 504 (1984)

\bibitem{jen85}
R. V. Jensen and R. Shankar, Phys. Rev. Lett. {\bf 54}, 1879 (1985)

\bibitem{deu91}
J. M. Deutsch, Phys. Rev. A {\bf 43}, 2046 (1991)

\bibitem{sre96}
M. Srednicki, J. Phys. A: Math. Gen {\bf 29}, L75 (1996);
{\bf 32}, 1163 (1999)

\bibitem{tas98}
H. Tasaki, Phys. Rev. Lett. {\bf 80}, 1373 (1998)

\bibitem{das03}
N. Dass, S. Rama, and B. Sathiaplan, Int. J. Mod. Phys. {\bf 18}, 2947 (2003)

\bibitem{rei08}
P. Reimann, Phys. Rev. Lett. {\bf 101}, 190403 (2008)

\bibitem{lin09}
N. Linden, S. Popescu, A. J. Short, and A. Winter,
Phys. Rev. E {\bf 79}, 061103 (2009)

\bibitem{exp}
T. Kinoshita, T. Wenger, D. S. Weiss, Nature {\bf 440}, 900 (2006);
S. Hofferberth et al., {\em ibid} {\bf 449}, 324 (2007)

\bibitem{rig07}
M. Rigol, V. Dunjko, V. Yurovsky, and M. Olshanii, Phys. Rev. Lett. {\bf 98}, 050405 (2007);
C. Kollath, A. M. L\"auchli, and E. Altman, {\em ibid} {\bf 98}, 180601 (2007);
S. R. Manmana, S. Wessel, R. M. Noack, and A. Muramatsu, {\em ibid} {\bf 98}, 
210405 (2007)

\bibitem{cra08}
M. Cramer, C. M. Dawson, J. Eisert, and T. J. Osborne,
Phys. Rev. Lett {\bf 100}, 030602 (2008);
M. A. Cazalilla, {\em ibid} {\bf 97}, 156403 (2006);
T. Barthel and U. Schollw\"ock, {\em ibid} {\bf 100},
100601 (2008);
A. Fleisch et al., Phys. Rev. A {\bf 78}, 033608 (2008)

\bibitem{rue69}
D. Ruelle, Statistical Mechanics, Benjamin,  New York (1969).

\bibitem{hob71}
A. Hobson, {\em Concepts in Statistical Mechanics}, 
Gordon and Breach, New York 1971.

\bibitem{realobs}
A. Y. Khinchin, Ch. 3 in {\em Mathematical Foundations of Quantum Statistics},
Graylock Press, NY (1960);
E. P. Wigner, Am. J. Phys. {\bf 31}, 6 (1963);
N.G. van Kampen, Ch. XVII.7 in  {\em Stochastic Processes 
in Physics and Chemistry}, Elsevier, Amsterdam 1992;
A. Sugita, Nonlinear Phenom. Complex Syst. {\bf 10}, 192 (2007);
O. Penrose, Ch. 1 in {\em Foundations of Statistical Mechanics},
Pergamon, Oxford 1970;

\bibitem{lof}
A. Martin-L\"of, Statistical Mechanics and the Foundations of Thermodynamics,
Lecture Notes in Physics 101, Springer, Berlin  1979.

\bibitem{geo95}
H.-O. Georgii, J. Stat. Phys. {\bf 80}, 1341 (1995)

\bibitem{pop06} 
S. Popescu, A. J. Short, and A. Winter, Nature Physics {\bf 2}, 754 (2006);
quant-ph/0511225;

\bibitem{rei07}
P. Reimann, J. Stat. Phys. {\bf 132}, 921 (2008)

\bibitem{rei07a}
P. Reimann,  Phys. Rev. Lett. {\bf 99}, 160404 (2007)

\bibitem{ediff}
M. Wilkinson, Phys. Rev. A {\bf 41}, 4645 (1990);
M. Wilkinson and E. J. Austin, J. Phys. A: Math. Gen {\bf 28}, 2277 (1995);
A. Bulgac, G. D. Dang, and D. Kusnezov,
Phys. Rev. E {\bf 54}, 3468 (1996);
D. Cohen, Phys. Rev. Lett. {\bf 82}, 4951 (1999)

\bibitem{gol06}
S. Goldstein, J. L. Lebowitz, R. Tumulka, and N. Zanghi,
J. Stat. Phys. {\bf 125}, 1197 (2006)

\bibitem{wik}
``Chebyshev's inequality'' in {\em Wikipedia, the free encyclopedia}.

\bibitem{skl93}
L. Sklar, {\em Physics and Chance}, Cambridge University Press 1993.

\bibitem{lldiu}
Landau and Lifshitz, {\em Statistical Physics},
Pergamon, Oxford 1980;
B. Diu, C. Guthmann, D. Lederer, and B. Roulet, 
{\em Elements de Physique Statistique}, Hermann, 
Paris 1996.

\bibitem{arg}
See Ref. \cite{sre96}, M. Feingold and A. Peres, 
Phys. Rev. A {\bf 34}, 591 (1986), and references 
therein.

\bibitem{sre94} 
M. Srednicki, Phys. Rev. E. {\bf 50}, 888 (1994);  
cond-mat/9410046

\bibitem{fei84}
M. Feingold, N. Moiseyev, and A. Peres, 
Phys. Rev. A {\bf 30}, 509 (1984)

\bibitem{wei92}
S. Weigert, Physica D {\bf 56}, 107 (1992);
B. Sutherland, Ch. 2.1 in {\em Beautiful Models}, 
World Scientific 2004;
A. Enciso and D. Peralta-Salas, Theor. Math. Phys. {\bf 148}, 1086 (2006)

\bibitem{rou09}
S. Roux, arXiv:0909.4620v1 [cond-mat.quant-gas]

\bibitem{f1}
In doing so, we have without loss of generality
assumed a labeling of the $E_n(\lambda)$ so that
$E_{n+1}(\lambda)\geq E_{n}(\lambda)$ for all $n$
(cf. (\ref{-1})) independently of $\lambda$, 
and thus local averages always involve the same 
subsets of levels in the vicinity of a given 
$E_n(\lambda)$, independently of $\lambda$,
see also Appendix E.

\bibitem{gol06a}
S. Goldstein, J. L. Lebowitz, R. Tumulka, and N. Zanghi,
Phys. Rev. Lett. {\bf 96}, 050403 (2006)

\bibitem{bar09}
C. Bartsch and J. Gemmer, Phys. Rev. Lett. {\bf 102}, 110403 (2009)

\bibitem{bun08}
L. A. Bunimovich, Nonlinearity {\bf 21}, T13 (2008) 
and references therein.

\bibitem{pec83}
P. Pechukas, Phys. Rev. Lett. {\bf 51}, 943 (1983);
T. Yukawa, Phys. Rev. Lett. {\bf 54}, 1883 (1985);
F. Haake, Quantum Signatures of Chaos, Springer, Berlin 2001.

\end{thebibliography}
\end{document}